\documentclass[aps,floatfix,twocolumn,byrevtex,superscriptaddress]{revtex4-1}
\usepackage[english]{babel}
\usepackage{amsmath}
\usepackage{graphicx}
\usepackage{float}
\usepackage{bm}
\usepackage{multirow}
\usepackage{dcolumn}
\usepackage{hyperref}
\usepackage[normalem]{ulem}
\usepackage[dvipsnames,usenames]{color}
\usepackage{xcolor}
\usepackage{soul}
\usepackage{array}

\begin{document}

\title{Muon spin rotation and infrared spectroscopy study of Ba\textsubscript{1-x}Na\textsubscript{x}Fe\textsubscript{2}As\textsubscript{2} }

\author{E.~Sheveleva}
\email[]{evgeniia.sheveleva@unifr.ch}
\author{B.~Xu}
\author{P.~Marsik}
\author{F.~Lyzwa}
\affiliation{University of Fribourg, Department of Physics and Fribourg Center for Nanomaterials, Chemin du Mus\'{e}e 3, CH-1700 Fribourg, Switzerland}

\author{B.~P.~P.~Mallett}
\affiliation{The MacDiarmid Institute for Advanced Materials and Nanotechnology and The~Dodd-Walls Centre for Photonic and Quantum Technologies, The University of Auckland, NZ-1010 Auckland, New Zealand}

\author{K.~Willa}
\author{C.~Meingast}
\author{Th.~Wolf}
\affiliation{Institute for Quantum Materials and Technologies - IQMT, Postfach 3640, DE-76021 Karlsruhe, Germany}

\author{T.~Shevtsova}
\author{Yu.~G.~Pashkevich}
\affiliation{O. O. Galkin Donetsk Institute for Physics and Engineering NAS of Ukraine, UA-03680 Kyiv, Ukraine}

\author{C.~Bernhard}
\email[]{christian.bernhard@unifr.ch}
\affiliation{University of Fribourg, Department of Physics and Fribourg Center for Nanomaterials, Chemin du Mus\'{e}e 3, CH-1700 Fribourg, Switzerland}

\date{\today}

\begin{abstract}
The magnetic and superconducting properties of a series of underdoped Ba\textsubscript{1-x}Na\textsubscript{x}Fe\textsubscript{2}As\textsubscript{2} (BNFA) single crystals with \( 0.19 \leq x\leq 0.34 \) has been investigated with the complementary muon-spin-rotation (\(\mu\)SR) and infrared spectroscopy techniques. 
The  focus has been on the different antiferromagnetic states in the underdoped regime and their competition with superconductivity, especially for the ones with a  tetragonal crystal structure and a so-called double-\textbf{Q} magnetic order. 
Besides the collinear state with a spatially inhomogeneous spin-charge-density wave (i-SCDW) order at \(x=0.24\) and 0.26, that was previously identified in BNFA, we obtained evidence for an orthomagnetic state with a “hedgehog”-type spin vortex crystal (SVC) structure at \(x=0.32\) and 0.34. 
Whereas in the former i-SCDW state the infrared spectra show no sign of a superconducting response down to the lowest measured temperature of about 10K, in the SVC state there is a strong superconducting response similar to the one at optimum doping. 
The magnetic order is strongly suppressed here in the superconducting state and at \(x=0.34\) there is even a partial re-entrance into a paramagnetic state at \(T<<T_c\).

\end{abstract}

\maketitle

\section{Introduction}

\noindent The phase diagram of the iron arsenide superconductors is characterized by a close proximity of the antiferromagnetic (AF) and superconducting (SC) orders~\cite{Paglione, BasovNature2011}. 
This is exemplified by the prototypical system BaFe\textsubscript{2}As\textsubscript{2} (Ba-122) for which large, high-quality single crystals are readily available. 
The undoped parent compound is an itinerant antiferromagnet with a Neel temperature of $T^{N} \approx 135~K${}~\cite{Paglione}. 
Upon electron- or hole doping in Ba(Fe\textsubscript{1-x}Co\textsubscript{x})\textsubscript{2}As\textsubscript{2} (BFCA) \cite{SefatPRL2008} or Ba\textsubscript{1-x}K\textsubscript{x}Fe\textsubscript{2}As\textsubscript{2} (BKFA) \cite{RotterPRL2008} and Ba\textsubscript{1-x}Na\textsubscript{x}Fe\textsubscript{2}As\textsubscript{2} (BNFA)~\cite{WangPRB2016, YiPRL2018}, the AF order gets gradually suppressed and superconductivity emerges well before the magnetic order vanishes. 
In this so-called underdoped regime the AF and SC orders coexist and compete for the same low-energy electronic states~\cite{Pratt, MarsikPRL2010, YiNComm2014, KimPRB2014}. 
Upon doping, the superconducting critical temperature, \(T_c\), and other SC parameters, like the magnetic penetration depth, \(\lambda\), or the condensation energy, \(\gamma_s\), are enhanced  whereas the AF order parameter (the staggered magnetisation) is reduced. 
The full suppression of the static AF order is observed around optimum doping at which \(T_c\), \(\lambda\), and \(\gamma_s\) reach their maximal values~\cite{StoreyPRB2013, HardyPRB2016}.
A further increase of the doping leads to a decrease of \(T_c\) in the so-called overdoped regime for which the AF spin fluctuations also diminish. 
This characteristic doping phase diagram is one of the reasons, besides the unconventional s$\pm$ symmetry of the SC order parameter, why AF fluctuations are believed to be responsible for the SC pairing~\cite{Paglione}.
Nevertheless, there exist other candidates for the SC pairing mechanism such as the nematic/orbital fluctuations~\cite{SaitoPRBR2011, FernandesNPhys2014}. Even a phonon mediated pairing or a coupled spin-phonon mechanism is not excluded yet~\cite{YndurainPRB2009, BingPRL2019}.

There is also a strong coupling between the spin, orbital, and lattice degrees of freedom that is exemplified by the coupled AF and structural phase transition from a tetragonal paramagnetic state with \(C_4\) symmetry at high temperature to an orthorhombic antiferromagnetic AF (o-AF) state with \(C_2\) symmetry~\cite{BoeriPRB2010, Egami2010, CohPRB2016}.
For this o-AF state, which occupies major parts of the magnetic phase diagram, the spins are antiparallel along \((0;\pi)\) and parallel along \((0;\pi)\), giving rise to a so-called single-\textbf{Q} or stripe-like AF order~\cite{CruzNature2008}.
Deviations from this o-AF order occur closer to optimum doping. 
For example, in BFCA the o-AF order and the associated lattice distortions are reported to become incommensurate and to be strongly suppressed by SC and eventually vanish below \(T_c\)~\cite{PrattPRL2011}.

A different type of AF order, for which the lattice structure remains tetragonal (\(C_4\) symmetry), albeit with a fourfold enlarged unit cell, was recently observed in the hole-doped BKFA and BNFA systems~\cite{AvciNComm2014, WasserPRBR2015, BohmerNComm2015, AllredPRB2015, MallettPRL2015, MallettEPL2015, AllredNPhys2016}. 
This tetragonal antiferromagnetic (t-AF) order can be described in terms of a so-called double-\textbf{Q} order due to a superposition of the single-\textbf{Q} states along \((0;\pi)\) and \((\pi;0)\).
It can be realized either with a non-collinear magnetisation of the single-\textbf{Q} components, corresponding to a so-called orthomagnetic or “spin-vortex-crystal” (SVC) order, or with a collinear magnetisation that gives rise to an inhomogeneous state for which the Fe magnetic moment either vanishes or is doubled~\cite{LorenzanaPRL2008, KhalyavinPRB2014, KangPRBR2015}.
The latter state is accompanied by a subordinate charge density wave, forming a so-called spin-charge-density wave (SCDW)~\cite{MallettPRL2015, FernandesPRB2016}.
Experiments on BNFA~\cite{MallettEPL2015} and BKFA~\cite{AllredPRB2015} have identified the SCDW order with the spins oriented along the c-axis direction~\cite{WasserPRBR2015}, suggesting that spin-orbit interaction plays an important role~\cite{ChristensenPRB2015}.
It is still unknown which factors are most relevant for stabilizing these single-\textbf{Q} and double-\textbf{Q} AF orders, and even an important role of disorder has been proposed~\cite{HoyerPRB2016}.
In this context, it is interesting that a “hedgehog”-type orthomagnetic state has recently been identified in underdoped KCa(Fe\textsubscript{1-x}Ni\textsubscript{x})\textsubscript{4}As\textsubscript{4} for which the K\textsuperscript{+} and Ca\textsuperscript{2+} ions reside in separate layers that alternate along the c-axis. 
It has been speculated that the SVC order is stabilized here by the broken glide symmetry across the FeAs planes or by a reduced cation disorder~\cite{MeierQMat2018, DingPRL2018}. 
Of equal interest is the recent observation of yet another magnetic phase in BNFA that occurs at \( 0.3 < x < 0.37 \), i.e. between the i-SCDW phase and optimum doping~\cite{WangPRB2016}.
The latter is accompanied by a tiny orthorhombic distortion and therefore has been discussed in terms of an o-AF order with a very small magnetic moment~\cite{WangPRB2016}. 
Alternatively, it could be explained in terms of one of the SVC phases with tetragonal (\(C_4\)) symmetry that is somewhat distorted or coexists with a small fraction of the o-AF phase.  

The above described questions have motivated us to further explore the complex magnetic phase diagram of the iron arsenides and its relationship with SC. 
Here, we present an experimental approach using the complementary techniques of muon spin rotation (\(\mu\)SR) and infrared spectroscopy to study a series of BNFA single crystals that span the underdoped regime with its various magnetic phases. 
In particular, we provide evidence that the recently discovered AF phase that occurs shortly before optimum doping likely corresponds to an orthomagnetic “hedgehog”-type SVC order. 

This paper is organized as follows. 
The experimental methods are presented in Section II. 
Subsequently, we discuss in Section III the \(\mu\)SR data and in Section IV the infrared spectroscopy data. 
We conclude with a discussion and summary in Section V. 

\section{Experimental methods}

\begin{figure}[htb!]
\includegraphics[width=1\columnwidth]{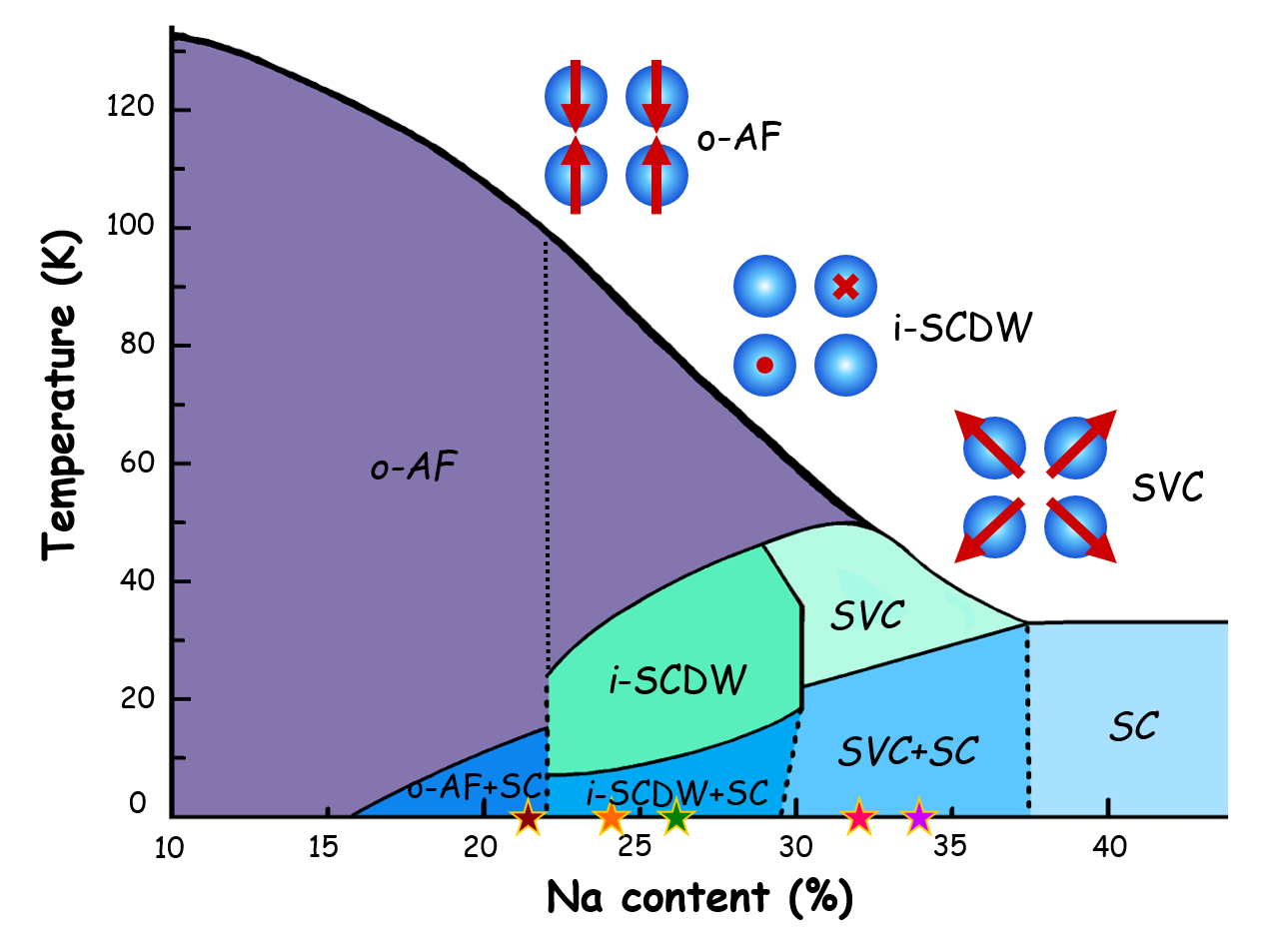}
\caption{(color online) Schematic phase diagram of Ba\textsubscript{1-x}Na\textsubscript{x}Fe\textsubscript{2}As\textsubscript{2} (BNFA) showing the different antiferromagnetic and superconducting phases and the location of the studied samples.}
\label{Fig1}
\end{figure}

Ba\textsubscript{1-x}Na\textsubscript{x}Fe\textsubscript{2}As\textsubscript{2} (BNFA) single crystals were grown in alumina crucibles with an FeAs flux as described in Ref.~\cite{WangPRB2016}.
They were millimeter-sized and cleavable yielding flat and shiny surface suitable for optical measurements. 
Selected crystals were characterized by x-ray diffraction refinement. 
For each crystal presented here, the Na-content, \(x\), was determined with electron dispersion spectroscopy with an accuracy of about \(\pm\)0.02 (estimated from the variation over the crystal surface). 
Figure~\ref{Fig1} shows the location of these crystals in the temperature vs doping phase diagram (marked with stars) that has been adopted from Ref.~\cite{WangPRB2016}. 
It also shows sketches of the various o-AF, i-SCDW and SVC magnetic orders. 
The magnetic and superconducting transition temperatures of the crystals, or of corresponding crystals from the same growth batch, have been derived from transport and from thermal expansion and thermodynamic experiments as described e.g. in Ref.~\cite{WangPRB2016}. 
Except for the SC transition of the crystals in the SCDW state at \(x=0.24\) and 0.26, the various magnetic and superconducting transitions have been confirmed with the \(\mu\)SR and infrared spectroscopy measurements as described below.
The bulk SC transition of the crystal at \(x=0.24\) is evident from additional specific heat data that are also shown below.

\begin{figure}[htb!]
\includegraphics[width=0.75\columnwidth]{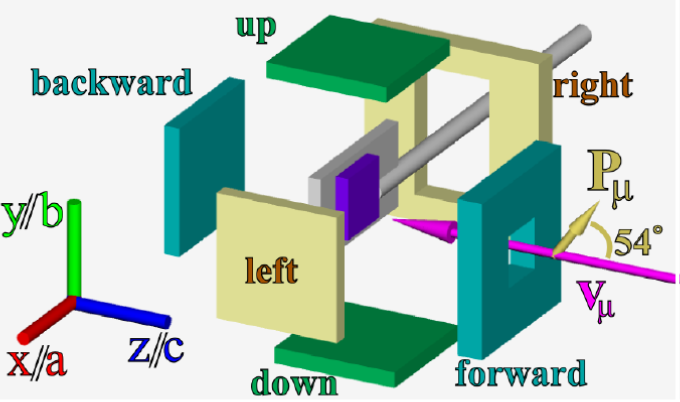}
\caption{(color online) Sketch of the geometry of the \(\mu\)SR setup at the GPS beamline of PSI showing the three pairs of position counters. The sample, shown in purple, has its c-axis aligned with the z-axis and the incoming muon beam. In the so-called transverse-field (TF) geometry, the muon spin is rotated by 54\(\textsuperscript{o}\) with respect to the z-axis. The external field for the transverse field experiments, \(\textbf{B}\textsubscript{ext}\), is applied parallel to the z-axis.}
\label{Fig2}
\end{figure}

The \(\mu\)SR experiments were performed at the general purposes spectrometer (GPS) at the \(\pi\)M3 beamline of the Paul Scherrer Institute (PSI) in Villigen, Switzerland which provides a beam of 100\% spin-polarized, positive muons. 
This muon beam was implanted in the crystals along the c-axis with an energy of about 4.2. MeV. 
These muons thermalize very rapidly without a significant loss of their initial spin polarization and stop at interstitial lattice sites with a depth distribution of about 100-200 \(\mu\)m. 
The magnetic and superconducting properties probed by the muons are thus representative of the bulk. 
The muons sites are assumed to be the same as in BKFA with a majority and a minority site that account for about 80\% and 20\% of the muons, respectively. 
As discussed in the Appendix A and also shown in Fig. 3(a) of Ref.~\cite{MallettEPL2015}, the majority site has a rather high local symmetry and is located on the line that connects the Ba and As ions along the c-axis (at the \((0,0,0.191)\) coordinate of the I4/mmm setting~\cite{ChristensenPRB2015}).
The minority site is located at \((0.4,0.5,0)\) and has a similar high local symmetry with the same direction and qualitative changes of the local magnetic field. 
The spin of the muons precesses in its local magnetic field, \(\textbf{B}_\mu\), with a frequency \(\nu_\mu=(\gamma_\mu/2\pi)\cdot B_\mu\), where \(\gamma_\mu=2\pi\cdot 135.5 MHz/T\) is the gyromagnetic ratio of the muon.
In a \(\mu\)SR experiment one measures the time evolution of the spin polarisation of an ensemble of (typically several million) muons, \(P(t)\). 
This is done via the detection of the asymmetry of the positrons that originate from the radioactive decay of the muons with a mean life time of \(\tau_\mu \approx 2.2 \mu s\) and which are preferentially emitted in the direction of P(t) at the instant of decay. 
This asymmetry is recorded within a time window of about 10\textsuperscript{-6} – 10\textsuperscript{-9}~s which allows one to detect magnetic fields ranging from about 0.1 Gauss to several Tesla. 
Most of the zero-field (ZF) and transverse field (TF) experiments reported here were performed in the TF-geometry using the so-called upward (u) and downward (d) counters which have higher and more balanced count rates than the forward (f) and backward (b) counters. 
The signal of the pair of fb-counters was only used in combination with the one from the ud-counters for the determination of the direction of \(\textbf{B}_\mu\) (as specified in the relevant figures).
The initial asymmetry of the ud-counter in the so-called transverse-field (TF) geometry, for which the muon spin is rotated by about 54\(\textsuperscript{o}\) (toward the upward counter) in the direction perpendicular to the momentum of the muon beam as shown in Fig.~\ref{Fig2}, is about 20-22\%. 
This variation of the initial asymmetry typically arises from a difference in the size and the exact positioning of the samples with respect to the positron counters as well as the so-called veto counter that is used for small samples to reduce the background signal due to muons that missed the sample. 
Further details about the \(\mu\)SR technique can be found e.g. in Refs.~\cite{Schenkmusr, Leemusr, Brewermusr}. 

The optical response was measured in terms of an in-plane reflectivity function \(R(\omega)\) at a near-normal angle of incidence with a Fourier-Transform Infrared (FTIR) spectrometer Bruker Vertex 70V in the frequency range from 40-8000~cm\textsuperscript{-1} with an \textit{in situ} gold evaporation technique~\cite{HomesAO1993}. 
Data were collected at different temperatures between 10~K and 300~K using a ARS Helitran cryostat. 
Room temperature spectra of the complex dielectric function in the near-infrared to ultra-violet (NIR-UV) range of 4000-52000~cm\textsuperscript{-1} were obtained with a commercial Woollam VASE ellipsometer.
The combined ellipsometry and reflectivity spectra were used to perform a Kramers-Kronig analysis to derive the complex optical response functions~\cite{Dressel2002} which in the following are expressed in terms of the complex optical conductivity \(\sigma(\omega)=\sigma\textsubscript{1}(\omega)+i\sigma\textsubscript{2}(\omega)\), or, likewise, the complex dielectric function, \(\epsilon(\omega)=\epsilon\textsubscript{1}(\omega)+i\epsilon\textsubscript{2}(\omega)\)), that are related according to \(\sigma(\omega)=i\frac{2\pi}{Z\textsubscript{0}}\omega\epsilon(\omega)\).
Below 40~cm\textsuperscript{-1} we extrapolated the reflectivity data with a Hagen-Rubens model \(R(\omega)=1-A\sqrt{\omega}\) in the normal state and a superconducting model \(R(\omega)=1-A\omega\textsuperscript{4}\) below \(T_c\).
On the high-frequency side above 52000~cm\textsuperscript{-1} we assumed a constant reflectivity up to 225000~cm\textsuperscript{-1} that is followed by a free-electron (\(\omega\textsuperscript{-4}\)) response.

\section{Muon Spin Rotation - \(\mu\)SR}

\begin{figure}[htb!]
\includegraphics[width=1\columnwidth]{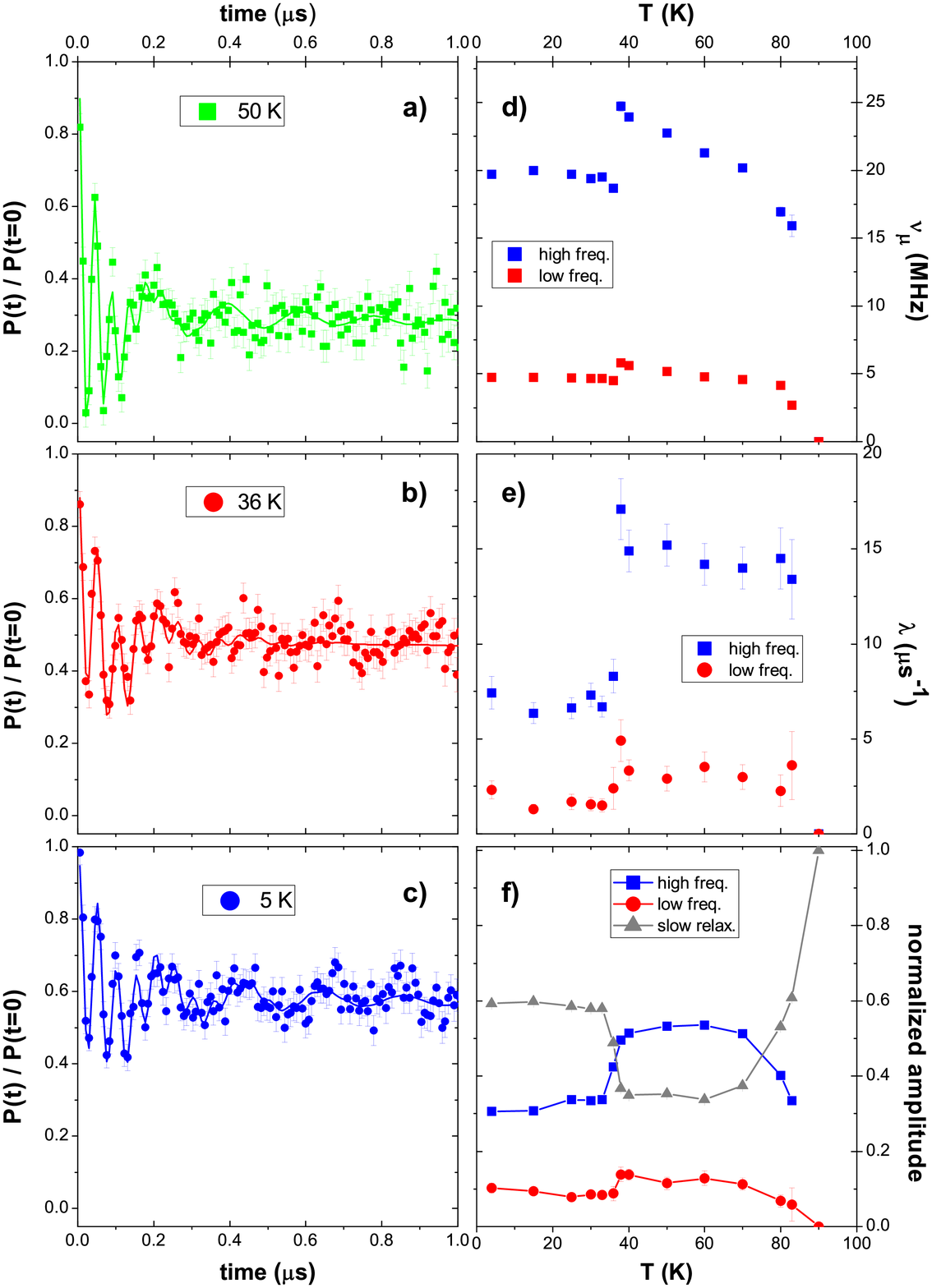}
\caption{(color online) Zero-field (ZF) \(\mu\)SR data of the BNFA crystal with \(x\approx0.24\) and \(T\textsubscript{c}\approx12K\) showing two magnetic transitions to an o-AF state below \(T\textsuperscript{N,1}\approx85K\) and the i-SCDW state below \(T\textsuperscript{N,1}\approx38K\). \textbf{(a)-(c)} ZF-\(\mu\)SR curves taken at 50K in the o-AF state and at 36K and 5K in the i-SCDW state, respectively. \textbf{(d)} and \textbf{(e)} Temperature dependence of the precession frequencies and relaxation rates of the oscillatory signals from two different muon sites, respectively. \textbf{(f)} Temperature dependence of the normalized amplitudes of the oscillatory signals and the slowly relaxing, non-oscillatory signal. Below \(T\textsuperscript{N}\), the latter arises mainly due to the non-orthogonal orientation of \(\textbf{B}_\mu\) and \(\textbf{P}\), apart from a small background due to muons that missed the sample.}
\label{Fig3}
\end{figure}

We start with the discussion of the zero-field (ZF)-\(\mu\)SR data for which only the internal magnetic moments contribute to the magnetic field at the muon site, \(\textbf{B}_\mu\). 

Figure~\ref{Fig3} summarizes the (ZF)-\(\mu\)SR study of the BNFA crystal with \(x\approx0.24\) that exhibits a transition from a high temperature paramagnetic state to an o-AF state at \(T\textsuperscript{N,1}\approx85K\) and a subsequent transition to a t-AF and i-SCDW state at \(T\textsuperscript{N,2}\approx38K\) that is followed by a SC transition at \(T\textsubscript{c}\approx12K\). 
Figures~\ref{Fig3}(a)-\ref{Fig3}(c) display characteristic, time-resolved spectra of the evolution of the muon spin polarisation, \(P(t)\), in the o-AF state at 50K and in the i-SCDW state at 36K and 5K. 
They exhibit clear oscillatory signals that are indicative of a bulk magnetic order. 
The solid lines show fits with the function:
\begin{equation}
	\label{eq:1}P(t)=P(0)\sum_{i=1}^2 A^{osc}_{i}\cos(\gamma_\mu B_\mu t+\Phi\textsubscript{i})e\textsuperscript{-\(\lambda_i t\)}+A^{non}_{3}e\textsuperscript{-\(\lambda_3 t\)},
\end{equation}
where \(A\textsubscript{i}\), \(B\textsubscript{\(\mu\),i}\), \(\Phi\textsubscript{i}\), and \(\lambda\textsubscript{i}\) account for the relative amplitudes of the signal, the local magnetic field at the muon sites, the initial phase of the muon spin, and the relaxation rates, respectively. 
The two oscillating signals with amplitudes \(A^{osc}_{1}\) and  \(A^{osc}_{2}\) arise from two muon sites with different local fields, as discussed in Ref.~\cite{MallettEPL2015}.
The non-oscillating signal  \(A^{non}_{3}\) results from the non-orthogonal orientation of \(\textbf{P}\) and \(\textbf{B}_\mu\). 
In addition, it contains a small contribution due to a non-magnetic background from muons that stopped outside the sample. 
The latter is typically less than 5\% of the total signal. 

The temperature dependence of the obtained fit parameters is displayed in Fig.~\ref{Fig3}(d) for the two precession frequencies, in Fig.~\ref{Fig3}(e) for the corresponding relaxation rates, and in Fig.~\ref{Fig3}(f) for the normalized amplitudes.
All three parameters exhibit pronounced changes at \(T\textsuperscript{N,2}\approx38K\). 
As outlined in Ref.~\cite{MallettEPL2015}, the decrease of the precession frequency, the relaxation rate and the amplitude of the oscillatory signal below \(T\textsuperscript{N,2}\approx38K\) are indicative of a transition from the o-AF to a t-AF and i-SCDW order. 
The only difference with respect to BKFA in Ref.~\cite{MallettEPL2015} is that the present BNFA crystal does not show any sign of a re-entrance towards an o-AF state below \(T\textsubscript{c}\), i.e. it remains in the i-SCDW state down to 5K without any noticeable anomaly at \(T\textsubscript{c}=12.3K\).  

\begin{figure}[htb!]
\includegraphics[width=1\columnwidth]{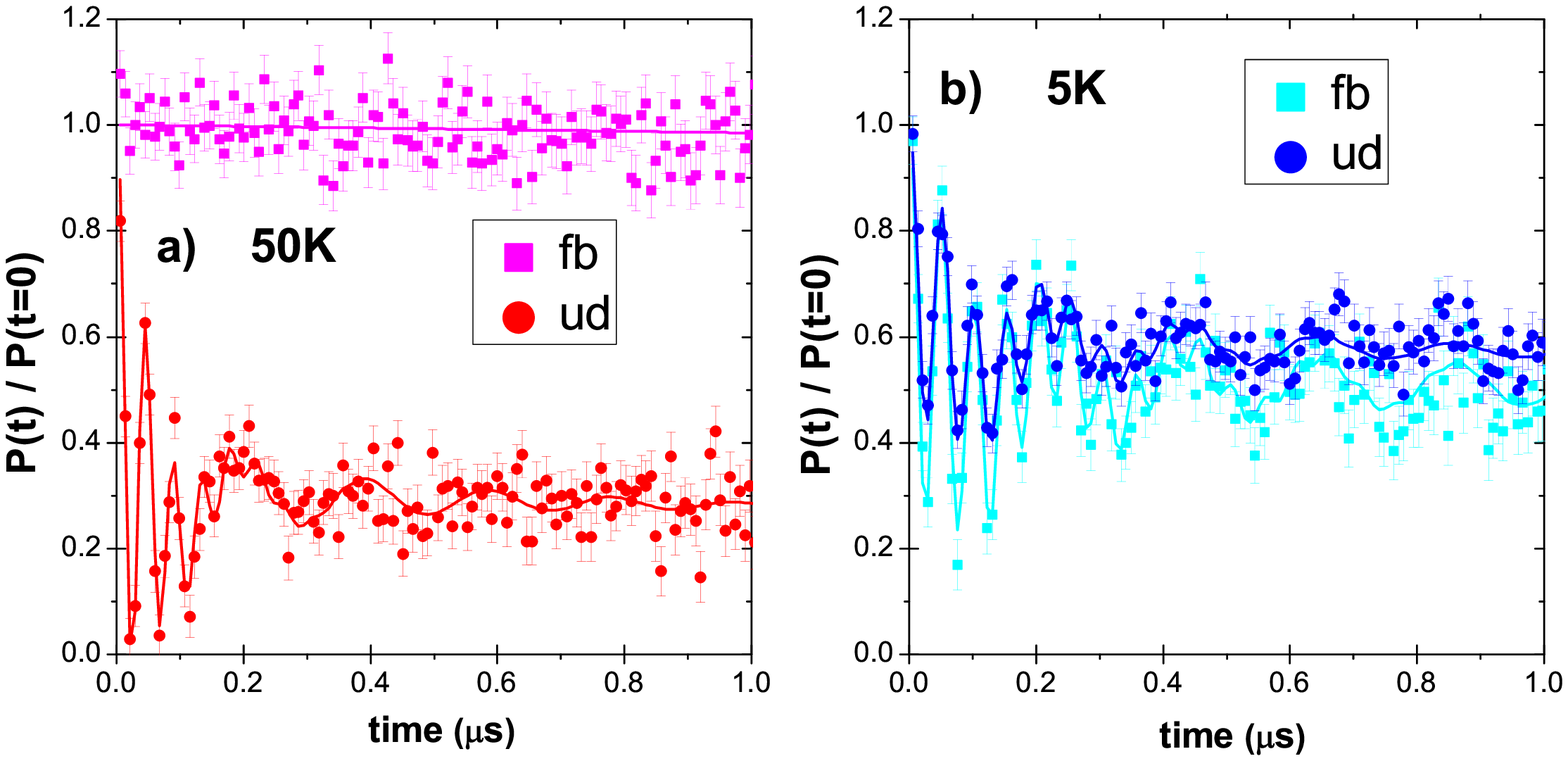}
\caption{(color online) ZF-\(\mu\)SR spectra of the \(x\approx0.24\) crystal showing the spin reorientation at the o-AF to i-SCDW transition. 
\textbf{(a)} ZF-\(\mu\)SR spectra of the pairs of forward-backward (fb) and up-down (ud) positron counters (see the sketch in Fig.~\ref{Fig2}) at 50K in the o-AF state. The absence of an oscillatory signal of the fb-counters confirms that \(\textbf{B}_\mu\) is parallel to the c-axis. 
\textbf{(b)} ZF-\(\mu\)SR spectra at 5K in the i-SCDW state for which the fb-counters show a large oscillatory signal suggesting an in-plane orientation of \(\textbf{B}_\mu\).}
\label{Fig4}
\end{figure}

Figures~\ref{Fig4}(a) and \ref{Fig4}(b) reveal that the ZF-\(\mu\)SR data show clear signatures of a change of the Fe-spin direction from an in-plane orientation in the o-AF phase to a c-axis orientation in the i-SCDW state. 
This is evident from the comparison of the amplitudes of the oscillatory signals of two different pairs of positron counters, i.e. of the upward (u) and downward (d) counters and the forward (f) and backward (b) counters (a sketch of the counter geometry is shown in Fig.~\ref{Fig2}). 
According to the calculations in Appendix B, the in-plane oriented Fe spins in the o-AF phase give rise to a local magnetic field at the muon site, \(\textbf{B}_\mu\), that is pointing along the c-axis, \(\textbf{B}_\mu//\textbf{c}\). 
This is because of the high symmetry of the majority muon site which is on a straight line between the As and Ba (or Na) ions. 
To the contrary, the c-axis oriented spins (on every second Fe site) in the i-SCDW phase cause \(\textbf{B}_\mu\) to be parallel to the ab-plane, \(\textbf{B}_\mu//\textbf{ab}\)~\cite{MallettEPL2015}.
From the sketch in Fig.~\ref{Fig2} it is seen that the former case with \(\textbf{B}_\mu//\textbf{c}\) (o-AF phase) gives rise to a vanishing oscillatory signal for the fb-counters and a large oscillatory signal for the ud-counters. 
Such a behavior is evident for the ZF-\(\mu\)SR spectra in the o-AF phase at 50K in Fig.~\ref{Fig4}(a). 
In contrast, for the ZF-spectra at 5K in the t-AF and i-SCDW state in Fig.~\ref{Fig4}(b), the fb-counters exhibit a large oscillatory signal that is characteristic of an in-plane orientation of \(\textbf{B}_\mu\) due to a c-axis orientation of the spins.  

Note that this change of the direction of \(\textbf{B}_\mu\) is also evident from the TF-\(\mu\)SR spectra (not shown) for which in the o-AF phase the applied field \(\textbf{B}\textsubscript{ext}\) and the field from the magnetic moments \(\textbf{B}\textsubscript{mag}\) are along the c-axis, yielding \(\textbf{B}_\mu=\textbf{B}\textsubscript{mag}\pm\textbf{B}\textsubscript{ext}\), whereas in the i-SCDW phase \(\textbf{B}\textsubscript{mag}\) is along the ab-plane such that \(\textbf{B}_\mu=\sqrt{\textbf{B}^2_{mag}+\textbf{B}^2_{ext}}\)~\cite{MallettEPL2015}.

\begin{figure}[htb!]
\includegraphics[width=0.95\columnwidth]{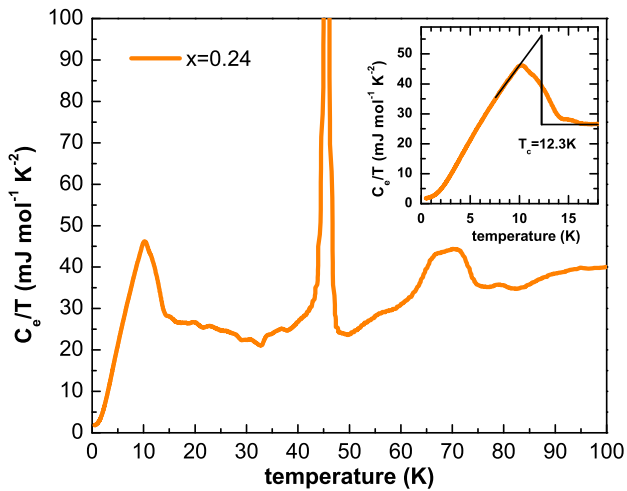}
\caption{(color online) Specific heat curve of a BNFA crystal at \(x\approx0.24\) with the phonon contribution subtracted as described in Ref.~\cite{WangPRB2016}. In addition to two sharp peaks due to magnetic transitions to the o-AF state below \(T^{N,1}\) and the i-SCDW state below \(T^{N,2}\), it exhibits a pronounced signature of bulk superconductivity. \textbf{Inset:} Magnification of the low temperature data from which the superconducting transition temperature of \(12.3K\) has been deduced using an entropy conserving construction (black line).}
\label{Fig5}
\end{figure}

Our \(\mu\)SR data thus provide clear evidence that the BNFA crystal with \(x\approx0.24\) undergoes a transition from a bulk o-AF state below \(T\textsuperscript{N,1}\approx85K\) with in-plane oriented spins to a bulk i-SCDW phase below \(T\textsuperscript{N,2}\approx38K\) that persists to the lowest measured temperature of 5K, even well below the SC transition at \(T\textsubscript{c}\approx12K\).
The bulk nature of the superconducting state with \(T_c\approx12K\) is evident from the specific heat data shown in Fig.~\ref{Fig5} for which the phonon contribution has been subtracted as described in Ref.~\cite{WangPRB2016}.
As detailed in the inset, the value of \(T_c=12.3K\) has been determined by the midpoint of the specific heat jump using an entropy conserving construction (black line).
The bulk nature of SC is evident from the more or less complete suppression of the electronic specific heat at very low temperature.
The specific heat curves also exhibit two more strong peaks at higher temperature that are due to the magnetic transitions into the o-AF state and the i-SCDW state at \(T^{N,1}\approx75K\) and \(T^{N,2}\approx45K\), respectively.
Note that these magnetic transition temperatures are somewhat lower than the ones obtained from the \(\mu\)SR data in Fig.~\ref{Fig4}.
Since the specific heat measurements have been performed on a smaller piece that was cleaved from the thick crystal measured with \(\mu\)SR (from the same side from which the infrared data have been obtained), the difference of the \(T^{N,1}\) and \(T^{N,2}\) values is most likely due to a variation of the Na content that is within the limits of \(\Delta x=\pm0.02\) as determined with electron dispersion spectroscopy.
Finally, note that the \(\mu\)SR data of the BNFA crystal with \(x\approx0.26\) (not shown) reveal a corresponding behavior as described above with two magnetic transitions from an o-AF phase below \(T\textsuperscript{N,1}\approx80K\) to a i-SCDW phase below \(T\textsuperscript{N,2}\approx42K\).

\begin{figure}[htb!]
\includegraphics[width=1\columnwidth]{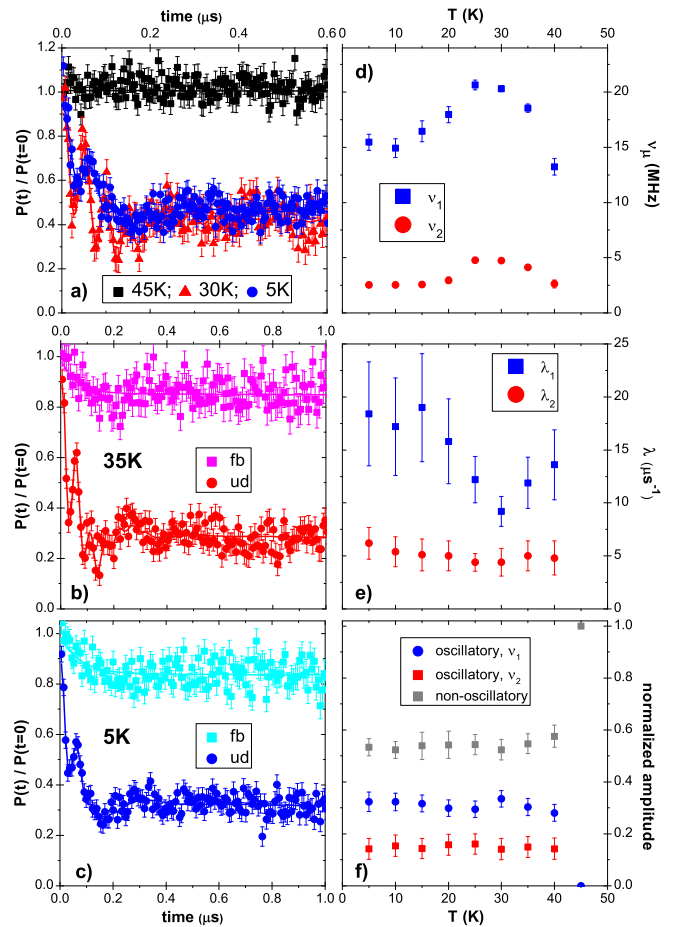}
\caption{(color online) ZF-\(\mu\)SR data at \(x\approx0.32\), \(T\textsubscript{c}\approx22K\). \textbf{(a)} ZF-\(\mu\)SR spectra showing the formation of a bulk magnetic state below \(T\textsuperscript{N}\approx45K\). \textbf{(b)} and \textbf{(c)} Comparison of the ZF spectra of the fb- and ud-counters that indicate a c-axis orientation of \(\textbf{B}_\mu\) and thus in-plane orientated spins. \textbf{(d)} to \textbf{(f)} Temperature dependence of the precession frequencies, \(\nu_1\) and \(\nu_2\), the relaxation rates, \(\lambda_1\) and \(\lambda_2\), and the normalized amplitudes of the oscillatory and the non-oscillatory signals, respectively.}
\label{Fig6}
\end{figure}

Next, we discuss the ZF-\(\mu\)SR data of the BNFA crystal with \(x\approx0.32\) and a Neel temperature of \(T\textsuperscript{N}\approx45K\) and \(T\textsubscript{c}\approx22K\). 
The nature of this AF order, which appears just shortly before static magnetism vanishes around optimal doping, remains to be identified. 
The thermal expansion measurements of Ref.~\cite{WangPRB2016} have shown that this AF order is accompanied by a very weak orthorhombic lattice distortion. 
In addition, magnetization measurements on Sr\textsubscript{1-x}Na\textsubscript{x}Fe\textsubscript{2}As\textsubscript{2} crystals with a corresponding magnetic order revealed an in-plane orientation of the magnetic moments~\cite{WangJPhys2019}.
Accordingly, these data have been interpreted in terms of an o-AF order with a very small magnetic moment~\cite{WangPRB2016} or a small magnetic volume fraction.
To the contrary, our ZF-\(\mu\)SR data in Fig.~\ref{Fig6} establish that this magnetic order is very strong and bulk-like (at least at \(T>T\textsubscript{c}\)).

Figure~\ref{Fig6}(a) confirms that the ZF-spectra below \(T\textsuperscript{N}\approx45K\) exhibit a large oscillatory signal with a rather high precession frequency.
The temperature dependence of the precession frequencies \(\nu_1\) and \(\nu_2\) (due to the two different muon sites) and of the corresponding amplitudes, as obtained from fitting with the function in eq.~(\ref{eq:1}), are displayed in Figs.~\ref{Fig6}(d) and \ref{Fig6}(f), respectively.
The value of \(\nu_1\) increases steeply below \(T\textsuperscript{N}\approx45K\) and reaches a maximum of \(\nu_\mu\approx22 MHz\) at 25 K. 
The latter is only about 10\% lower than the one obtained for the \(x\approx0.24\) sample in the o-AF state where it reaches a maximum of \(\nu_\mu\approx24.5 MHz\) at 40 K (see Fig.~\ref{Fig3}(d)). 
Note that the precession frequency is proportional to \(\textbf{B}_\mu\) and thus to the magnitude of the magnetic moment, given that the muon site remains the same (which is most likely the case). 
Figure~\ref{Fig6}(f) shows that the amplitude of the oscillatory signal amounts to about 60\% of the total amplitude. 
Assuming that the local field \(\textbf{B}_\mu\) is pointing along the c-axis and thus is at an angle of 54\(\textsuperscript{o}\) with respect to \(\textbf{P}\) (see Fig.~\ref{Fig2}), this suggests that the magnetic volume fraction is close to 100\%. 
The \(\mu\)SR data are therefore incompatible with an o-AF state that has a small magnetic moment or only a very small volume fraction.  

\begin{figure}[htb!]
\includegraphics[width=0.7\columnwidth]{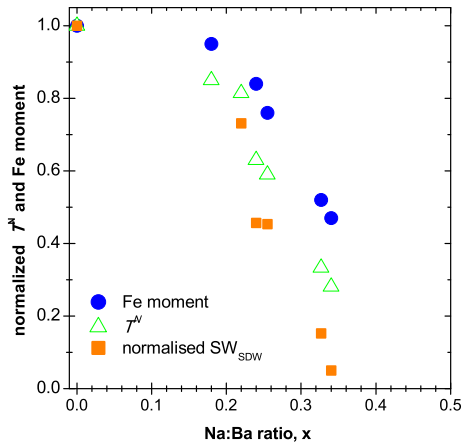}
\caption{(color online) Doping dependence of the normalized values (to the ones at x=0) of the Neel-temperature, \(T^N\), the magnetic Fe moment estimated from \(\mu\)SR (derived as described in the text) and the spectral weight of the SDW peak from IR spectroscopy for the o-AF phase at \(x<0.3\) and the suspected orthomagnetic SVC phase at \(x>0.3\).}
\label{Fig7}
\end{figure}

Instead, it seems likely that the observed magnetic state at \(x\approx0.32\) corresponds to the orthomagnetic, so-called “hedgehog”-type SVC order for which the spins are oriented within the ab-plane as shown in the Appendix B and sketched in Fig.~\ref{Fig1}.
The in-plane orientation of the magnetic moments is evident from the comparison of the ZF-\(\mu\)SR spectra of the forward-backward (fb) and up-down (ud) pairs of positron counters in Figs.~\ref{Fig6}(b) and \ref{Fig6}(c). 
Here the fb-signal has no detectable oscillatory component, meaning that \(\textbf{B}_\mu\) is mainly directed along the c-axis and the magnetic moments thus are oriented within the ab-plane. 
The small component (about 15\%) with a fast initial relaxation might arise from disordered regions. 
It might arise from a minority o-AF phase that would also explain the small orthorhombic distortion seen in Ref.~\cite{WangPRB2016}. 
Our \(\mu\)SR data are thus indicative of an SVC order at \(x=0.32\) as discussed in the Appendix B and sketched in Fig.~\ref{Fig1}.
According to the calculations in Appendix B, the value of \(\textbf{B}_\mu\) for the loop-type SVC order is zero whereas for the hedgehog-type SVC order it is about 40\% higher than for the single-\textbf{Q} AF order (assuming the same magnitude of the magnetic moment). 
When comparing the values of the precession frequencies with the one of the parent compound at x=0, with \(\nu_\mu\approx29 MHz\) and a magnetic moment of about 1 \(\mu_B\)/Fe ion as reported in Refs.~\cite{BernhardPRB2012, AczelPRB2008}, we thus obtain an estimate of the magnetic moment of the hedgehog SVC phase at \(x\approx0.32\) of about 0.5 \(\mu_B\)/Fe ion.
Likewise, the magnetic moment in the o-AF phase at \(x\approx0.24\) with \(\nu_\mu\approx24.5 MHz\) at 40K amounts to 0.85 \(\mu_B\)/Fe ion. 
Figure~\ref{Fig7} shows the resulting doping dependence of the estimated magnetic moment and the corresponding Neel temperature which both evolve continuously and tend to vanish around \(x=0.36-0.37\).

Notably, Fig.~\ref{Fig6}(d) reveals that the onset of SC at \(x\approx0.32\) is accompanied by a pronounced reduction of the precession frequency, from \(\nu_\mu\approx21 MHz\) at \(T\geq T_c\approx22K\) to \(\nu_\mu\approx15.5 MHz\) at \(T<<T_c\), and thus of the magnetic moment of the SVC order. 
Figure~\ref{Fig6}(e) shows that there is also a clear increase of the relaxation rate below \(T_c\approx22K\), which suggests that the magnetic order parameter becomes less homogeneous in the SC state. 
Nevertheless, the amplitude of the magnetic signal does not show any sign of a suppression below \(T_c\), suggesting that the magnetic order remains a bulk phenomenon even at \(T<<T_c\). 

\begin{figure}[htb!]
\includegraphics[width=1\columnwidth]{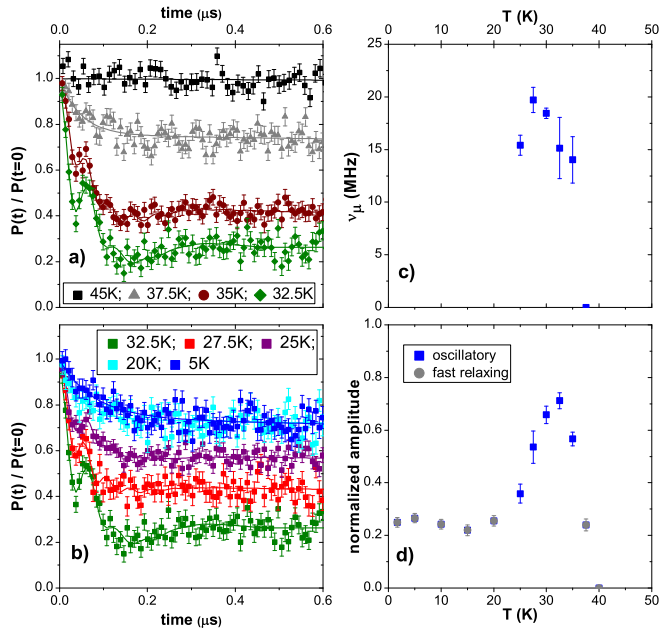}
\caption{(color online) ZF-\(\mu\)SR data at \(x\approx0.34\). \textbf{(a)} and \textbf{(b)} ZF-\(\mu\)SR spectra in the magnetic state below \(T\textsuperscript{N}\approx38K\) above and below \(T_c\approx25K\), respectively. \textbf{(c)} Temperature dependence of the precession frequency and \textbf{(d)} of the normalized amplitudes of the oscillatory signal and the non-oscillatory but fast relaxing signal that both arise from regions with large magnetic moments.}
\label{Fig8}
\end{figure}

Figure~\ref{Fig8} reveals that the suppression of the magnetic order due to the competition with SC becomes even more severe for the BNFA crystal with \(x\approx0.34\). 
The magnetic signal in the ZF-spectra develops here below \(T\textsuperscript{N}\approx38K\) and the frequency and amplitude of the precession signal are rising rapidly to values of \(\nu_\mu\approx19.5 MHz\) and about 65\%, respectively, at 30K. 
These are characteristic signatures of a bulk AF order that seems to be of the SVC type for the same reasons as discussed above for the \(x\approx0.32\) crystal. 
Notably, the onset of SC below \(T_c\approx30K\) at \(x\approx0.34\) gives rise to a much stronger suppression of the magnetic order than at \(x\approx0.32\). 
Not only the frequency is rapidly suppressed here but, as shown in Fig.~\ref{Fig8}(d), even the amplitude of the magnetic signal gets strongly reduced to about 25\% below 20K. 
This highlights that the magnetic order becomes spatially inhomogeneous with a large fraction of the sample re-entering a paramagnetic state. 
A similar re-entrance behaviour of the AF order was previously only observed for BFCA crystals in the region very close to optimum doping~\cite{Pratt}.

\begin{figure}[htb!]
\includegraphics[width=1\columnwidth]{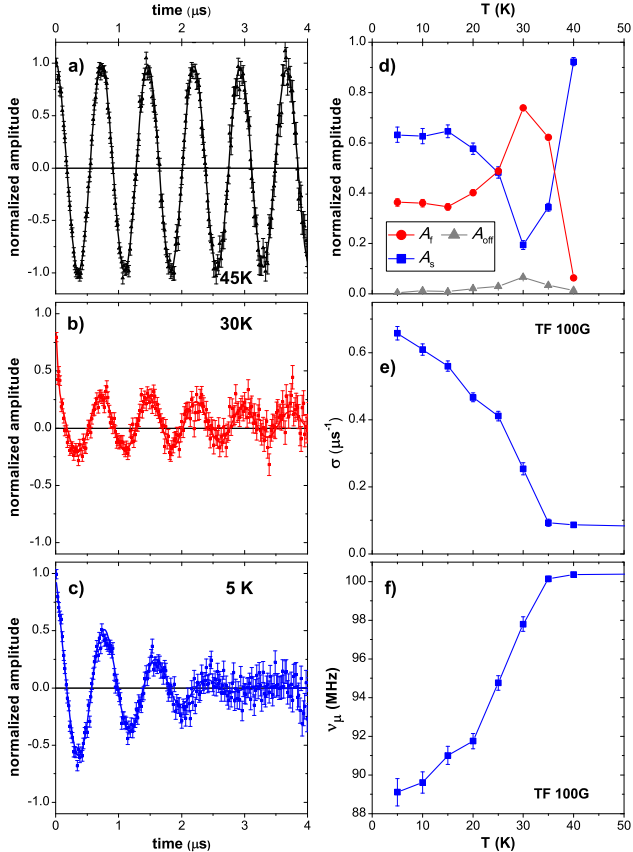}
\caption{(color online) TF-\(\mu\)SR spectra at 100G for \(x\approx0.34\). \textbf{(a)} - \textbf{(c)} TF-\(\mu\)SR spectra at \(T=45K>T^N\approx38K\), \(T^N>T=30K\geq T_c\approx30K\), and \(T=5K<<T_c\), respectively. \textbf{(d)} Temperature dependence of the normalized amplitudes (\(A_f\) and \(A\textsubscript{off}\)) of the magnetic signals and the non-magnetic signal (\(A_s\)) as described in the text. \textbf{(e)} and \textbf{(f)} Temperature dependence of the Gaussian relaxation rate, \(\sigma\), and the precession frequency, \(\nu_\mu\), of the non-magnetic signal (\(A_s\)) showing an enhanced relaxation and diamagnetic shift due to the superconducting vortex lattice below \(T_c\approx30K\).}
\label{Fig9}
\end{figure}

This re-entrance of large parts of the sample volume from a magnetic state at \(T\sim T_c\approx30K\) to a non-magnetic state at \(T<<T_c\approx30K\) is also evident from the 100G TF-\(\mu\)SR data in Fig.~\ref{Fig9}.
The solid lines in Figs.~\ref{Fig9}(a)-\ref{Fig9}(c) show fits with the function:
\begin{eqnarray}
\label{eq:2}P(t)=P(0) \{ A_{f} \cos \left(\gamma_{\mu} B_{\mu, f} t+\Phi_{f}\right) e^{-\lambda_{f} t} \nonumber \\ +A_{off} + A_{s} \cos \left(\gamma_{\mu} B_{\mu, s} t+\Phi_{s}\right) e^{-\frac{1}{2} \sigma^{2} t^{2}} \}.
\end{eqnarray}

The first two terms describe the magnetic signal.
The fast relaxing one with the normalized amplitude, \(A_f\), accounts for the strongly damped or even overdamped oscillatory part and the constant term with amplitude, \(A\textsubscript{off}\), for the non-oscillatory part of the magnetic signal that arises below \(T^N\).
The third term represents the non-magnetic signal with a Gaussian relaxation rate, \(\sigma\). 
The value of \(\sigma\) is much smaller than the one of \(\lambda_f\) and is governed above \(T_c\) by the nuclear spins and below \(T_c\) by the SC vortex lattice. 
The frequency of this non-magnetic signal is determined by the external magnetic field, except for the diamagnetic shift in the SC state. 
In contrast, the signal from the magnetic regions is governed by the internal magnetic moments (the contribution of the 100G TF is considerably smaller) which yield a higher frequency and a much faster relaxation such that this signal vanishing on a time scale of less than 0.5 \(\mu\)s. 

Figure~\ref{Fig9}(d) shows the temperature dependence of the normalized amplitudes of the magnetic signals \(A_f\) and \(A\textsubscript{off}\), and of the non-magnetic signal, \(A_s\).  
It reveals that the magnetic volume fraction increases rapidly to about 80\% at 30K and then decreases again below \(T_c\) to about 35\% at low temperature. 
Correspondingly, the non-magnetic fraction is reduced to about 20\% at 30K and increases again to about 65\% well below \(T_c\). 
Figures~\ref{Fig9}(e) and \ref{Fig9}(f) show that this non-magnetic part exhibits clear signs of a SC response in terms of a diamagnetic shift and an enhanced relaxation from the SC vortex lattice, respectively, that both develop below \(T_c\approx30K\).
From this Gaussian relaxation rate the value of the in-plane magnetic penetration depth, \(\lambda\textsubscript{ab}\), can be derived according to: \(\frac{\sigma}{1.23}\left[\mu s\right]=\frac{7.086\times 10\textsuperscript{-4}}{\lambda^2_{ab}}\left[nm\textsuperscript{-2}\right]\), as outlined, e.g., in Refs.~\cite{PumpinPRB1990, BernhardPRB1995}.
This yields a low temperature value of the magnetic penetration depth of \(\lambda\textsubscript{ab}(T\rightarrow0)\approx350 nm\) and for the related SC condensate density \(\frac{n_s}{m^{*}_{ab}}=\frac{1}{\mu_0 e^2\lambda^{2}_{ab}}\) of \(\frac{n_s}{m^{*}_{ab}}=2.7\cdot10^{20}\cdot\frac{m^{*}_{ab}}{m_e}\left[cm^{-3}\right]\), where \(\mu_0\), \(e\), \(m^{*}_{ab}\), \(m_e\) are the magnetic vacuum permeability, the elementary charge of the electron, and the effective band mass and bare mass of the electron, respectively.  
This value has to be viewed as an upper limit to the penetration depth (lower limit to the condensate density), since the Gaussian function is symmetric in frequency space and thus does not capture the asymmetric “lineshape” of the frequency distribution due to a vortex lattice which has a tail toward higher frequency. 
Also note that from these \(\mu\)SR data we cannot draw firm conclusions about the SC properties in the magnetic regions for which the relaxation due to the SC vortex lattice is much weaker than the magnetic one.

\section{Infrared Spectroscopy}

The \(\mu\)SR study of the magnetic and superconducting properties of the BNFA crystals presented in Section III has been complemented with infrared spectroscopy measurements as shown in the following.

\begin{figure*}[htb!]
\includegraphics[width=1.05\textwidth]{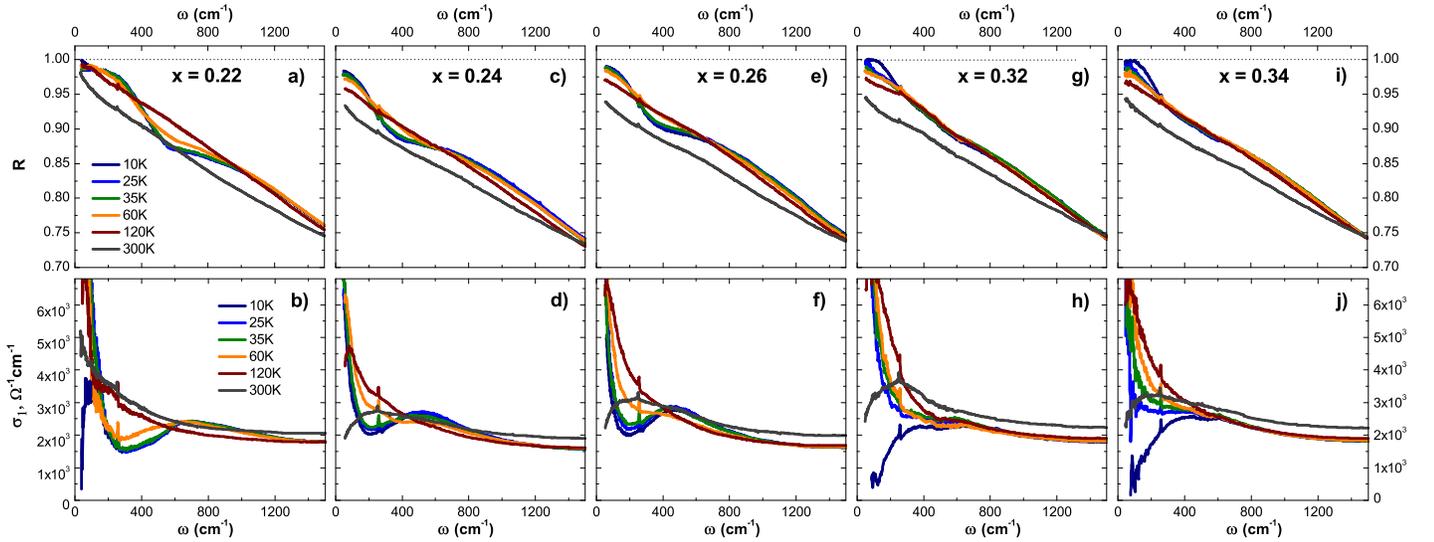}
\caption{(color online) Temperature dependent infrared optical response of BNFA crystals with \(0.22\leq x\leq0.34\). The upper panels show the reflectivity spectra at different temperatures in the paramagnetic state and in the various AF phases. The lower panels display the corresponding spectra of the real part of the optical conductivity obtained from a Kramers-Kronig analysis, as described in section II.}
\label{Fig10}
\end{figure*}

Figure~\ref{Fig10} gives an overview of the temperature dependent spectra of the measured reflectivity \(R(\omega)\) (upper panels) and of the obtained real part of the optical conductivity \(\sigma_1(\omega)\) (lower panels) for the crystals with \(x=0.22, 0.24, 0.26, 0.32\) and \(0.34\) that cover the different AF orders of the BNFA phase diagram in the normal and in the superconducting states (see Fig.~\ref{Fig1}).

\begin{figure*}[htb!]
\includegraphics[width=0.7\textwidth]{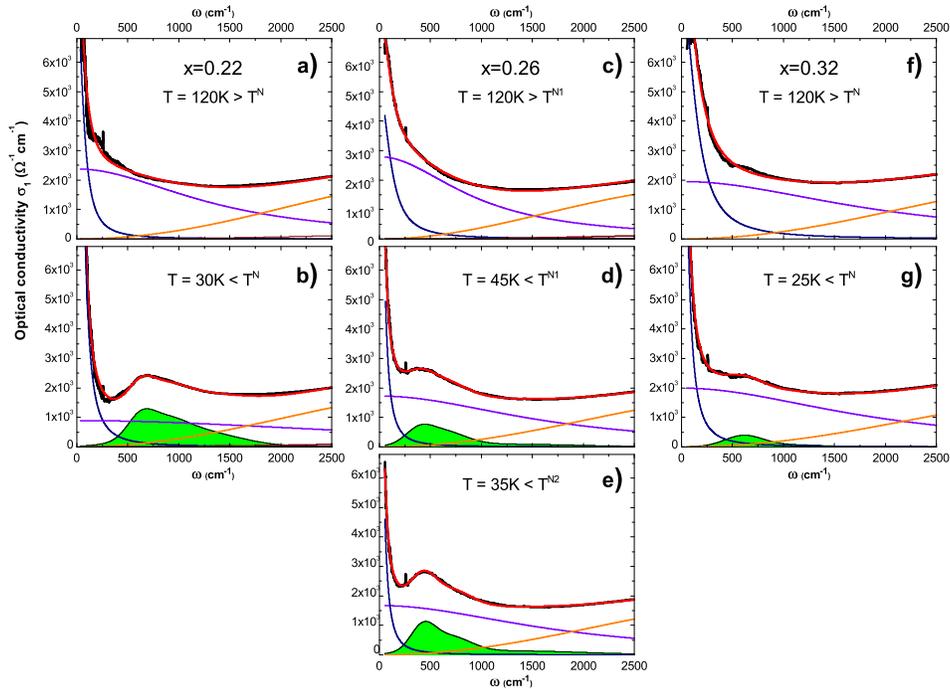}
\caption{(color online) Selected spectra of the optical conductivity in the paramagnetic and the different AF states and their fitting using the model described in equation~(\ref{eq:3}). The upper panels show the spectra in the paramagnetic state as described by the sum of a narrow and a broad Drude-peak (dark and light blue lines) and a Lorentz oscillator that account for the  free carriers and  the low-energy interband transitions, respectively. The lower panels display the spectra in the various AF states, i.e. at \(x=0.22\) in the o-AF state (left panel), at \(x=0.26\) in the intermediate o-AF and the i-SCDW states at low temperature, and at \(x=0.32\) in the suspected SVC phase. The green shaded area shows the pair-breaking peak that arises from the excitations across the SDW gap and has been accounted for with a sum of two Gaussian functions.}
\label{Fig11}
\end{figure*}

Figure~\ref{Fig11} displays representative spectra of the infrared conductivity in the paramagnetic and the various AF states and shows their fitting with a model function that consists of a sum of Drude-, Lorentz- and Gaussian oscillators:
\begin{equation}
\begin{aligned}
&\begin{array}{c}
\sigma_1(\omega)=\frac{2\pi}{Z_0}\left[\sum_{j}\frac{\omega^2_{pD_j}\gamma_{Dj}}{\omega^2+\gamma^2_{Dj}}+\sum_{k}\frac{\gamma_k\omega^2S^2_k}{(\omega^2_{0_k}-\omega^2)^2+\gamma^2_k\omega^2}\right] \\
+\sum_{i=1}^{3}S_{G_i}\cdot e\textsuperscript{\(-\frac{(\omega-\omega_{0G_i})^2}{2\gamma^2_{G_i}}\)}.
\label{eq:3}
\end{array}
\end{aligned}
\end{equation}
The first term contains two Drude peaks that account for the response of the itinerant carriers, each described by a plasma frequency, \(\omega_{pD_j}\), and a broadening \(\gamma_{Dj}\) that is proportional to the scattering rate, \(1/\tau_{D_j}\).
The Lorentz oscillators in the second term with an oscillator strength, \(S_k\), resonance frequency, \(\omega_{0_k}\), and linewidth \(\gamma_k\), describe the low energy interband transitions in the mid-infrared region that are typically weakly temperature dependent~\cite{MarsikPRBR2013}.
The Gaussian oscillators in the third term with the oscillator strength, \(S_{G_i}\), eigen-frequency, \(\omega_{0G_i}\), and linewidth, \(\gamma_{G_i}\), represent the so-called pair-breaking peak that develops in the itinerant AF state due to the excitation of the electronic quasi-particles across the gap of the spin density wave (SDW)~\cite{Dressel2002}. 
The spectral weight of this SDW peak originates from the Drude-bands whose spectral weight is accordingly reduced below \(T^N\). 
The sharp and much weaker feature around 260 cm\textsuperscript{-1} corresponds to an infrared-active phonon mode, the so-called Fe-As stretching mode~\cite{AkrapPRBR2009}, that has not been included in the modelling.

The upper row of panels in Fig.~\ref{Fig11} shows  the spectra in the paramagnetic state at \(T>T^N\) that are described by a sum of two Drude-bands, a broad and a narrow one, plus one Lorentzian oscillator. 
A similar model  was previously used to describe the spectra in the paramagnetic normal state of corresponding BKFA and BFCA crystals~\cite{WangPRB2016, WuPRB2009, BarisicPRB2010, KimPRB2010, HeumenEPL2010, DaiPRL2013, BingXuPRB2017, MallettPRB2017}.
Based on the comparison with the electronic scattering rate of Raman experiments in the so-called \(A_{1g}\) and \(B_{2g}\) scattering geometries of the incident and reflected laser beam~\cite{MuschlerPRBR2009}, which allow one to distinguish between the response of the hole-like and electron-like bands, we assign the broad Drude-peak in the infrared response to the hole-like bands near the center of the Brilluoin-zone (\(\Gamma\)-point) and the narrow Drude-peak to the electron-like bands near the boundary of the Brilluoin-zone (\(X\)-point), respectively. 

The lower panels of Fig.~\ref{Fig11} show corresponding spectra and their fitting for the different AF phases in the normal state above \(T_c\). 
The spectra are described by two additional Gaussian functions to account for the spectral weight that is accumulated above the SDW gap at \(\omega>2\Delta^{SDW}\). 
This so-called SDW peak arises from the quasi-particle excitations across the SDW gap. 
Its area (shaded in green) is a measure of the spectral weight of the itinerant charge carriers that contribute to the staggered magnetic moment of the SDW and thus is representative of the magnitude of the AF order parameter, see e.g. Figs.~3 and 8 of Ref.~\cite{MallettPRB2017}.

\begin{figure}[htb!]
\includegraphics[width=1\columnwidth]{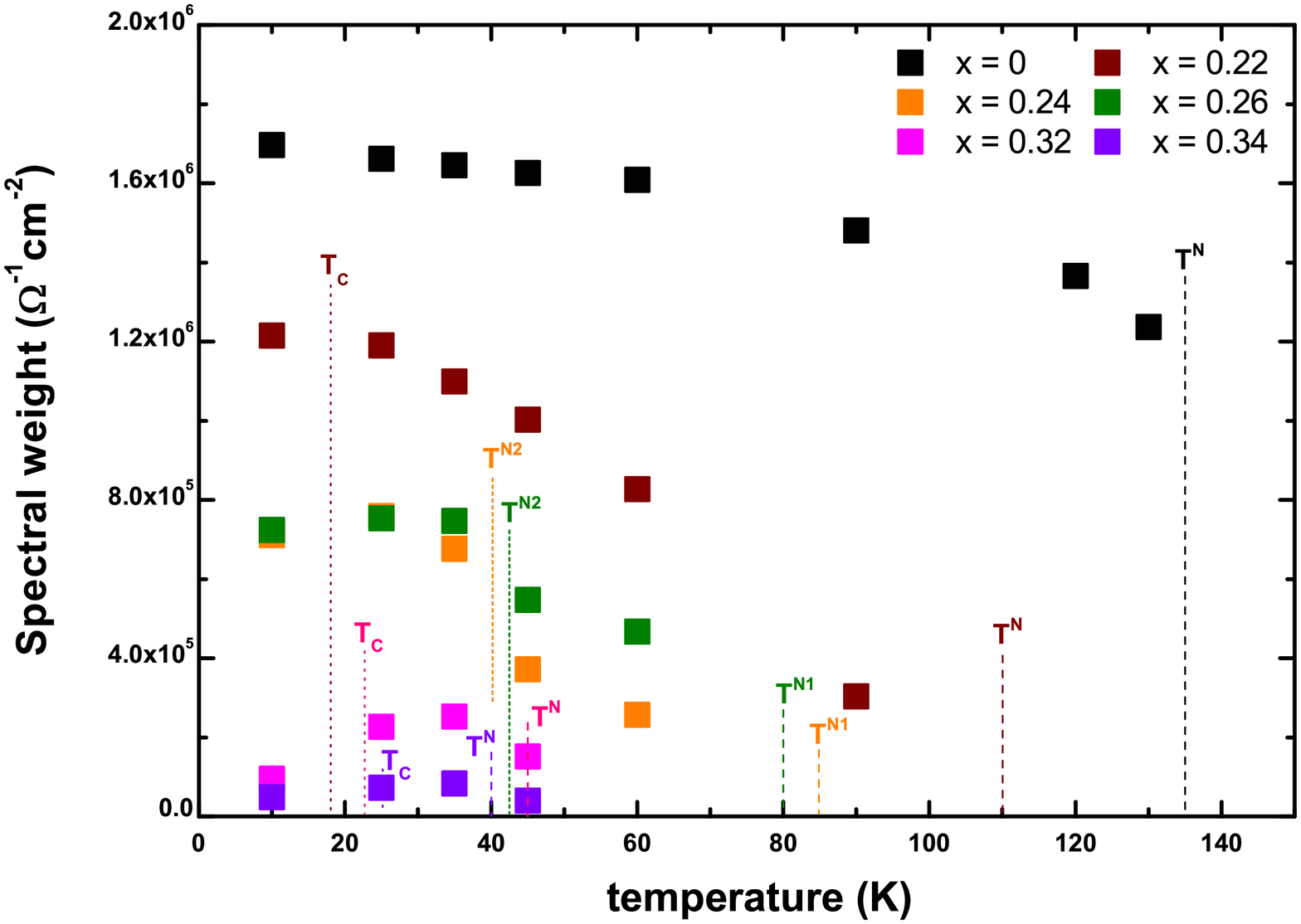}
\caption{(color online) Temperature evolution of spectral weight of the SDW peak. Arrows mark the transitions into the o-AF state below \(T^N\) at \(x=0\) and \(0.22\), into the successive o-AF and i-SCDW states below \(T^{N,1}\) and \(T^{N,2}\), respectively at \(x=0.24\) and \(0.26\), and into the SVC phase below \(T^N\) at \(x=0.32\) and \(0.34\).}
\label{Fig12}
\end{figure}

Figure~\ref{Fig12} gives a full account of the obtained temperature dependence of the spectral weight (SW) of the SDW peak for the series of BNFA crystals. 
Also shown, for comparison, are the corresponding data for the undoped parent compound at \(x=0\) that are adopted from Ref.~\cite{MallettPRB2017}.
The various AF and SC transition temperatures are marked with arrows in the color code of the experimental data. 
Figure~\ref{Fig12} reveals that the spectral weight of the SDW peak is continuously suppressed as a function of hole doping, \(x\). 
Concerning the temperature dependence, for the sample with \(x=0.22\), which exhibits an o-AF order and an orthorhombic (\(C_2\)) structure below \(T^N\approx110K\), the SW of the SDW grows continuously below \(T^N\), without any noticeable anomaly due to the competition with superconductivity below \(T_c\approx15K\).
For the samples with \(x=0.24\) and \(x=0.26\), the SW of the SDW peak exhibits a sudden, additional increase at the transition from the intermediate o-AF state with \(C_2\) symmetry below \(T^{N,1}\approx85K\) to the i-SCDW order with \(C_4\) symmetry below \(T^{N,2}\approx40K\).
A similar SW increase of the SDW peak in the i-SCDW state was previously reported in Ref.~\cite{MallettPRL2015} for a corresponding BKFA crystal. 
For the BNFA samples at \(x=0.24\) and \(0.26\) the infrared spectra show no sign of a bulk-like superconducting response down to the lowest measured temperature of 10K. 
Note, however, that a bulk SC transition with \(T_c=12.3K\) at \(x=0.24\) is evident from the specific heat data in Fig.~\ref{Fig5}.
Finally, at \(x=0.32\) and \(0.34\) the SDW peak acquires only a rather small amount of SW in the spin vortex crystal (SVC) state below \(T^N\approx45K\) and \(40K\), respectively, that is assigned based on the \(\mu\)SR data as discussed in section III. 
Nevertheless, as shown in Figs.~\ref{Fig10} and \ref{Fig12}, a SDW peak can still be identified in the infrared spectra. 
Moreover, pronounced anomalies occur in the SC state below \(T_c\approx20K\) and \(25K\), respectively, where the SW of the SDW peak is reduced. 
This SC-induced suppression of the SDW peak corroborates the \(\mu\)SR data which reveal a corresponding suppression of the ordered magnetic moment at \(x=0.32\) and of the magnetic volume fraction at \(x=0.34\) (see Figs.~\ref{Fig6} and \ref{Fig8}, respectively).

\begin{figure}[htb!]
\includegraphics[width=0.7\columnwidth]{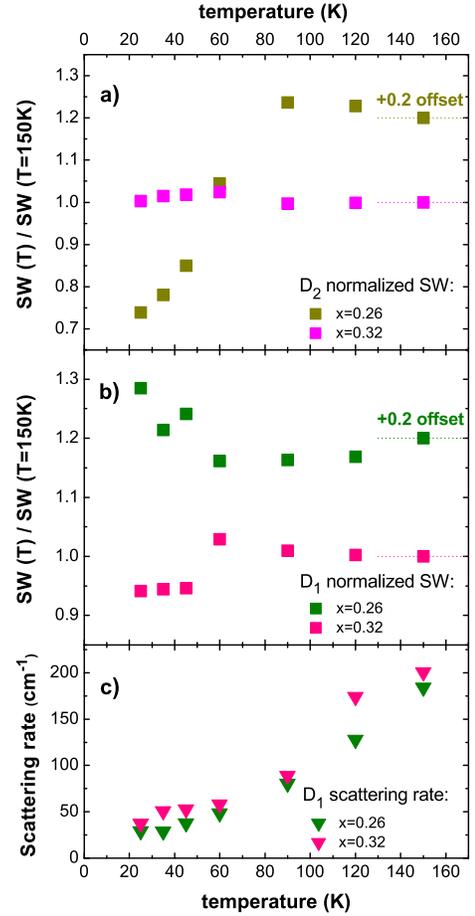}
\caption{(color online) Temperature dependence of the Drude parameters at \(x=0.26\) and \(x=0.32\). \textbf{(a)} and \textbf{(b)} Normalized spectral weight (with respect to the one at 150K) of the broad Drude peak (\(D_2\)) and the narrow Drude peak (\(D_1\)), respectively. \textbf{(c)} Temperature dependence of the scattering rate of the narrow Drude peak, \(\Gamma_{D_1}\).}
\label{Fig13}
\end{figure}

Figure~\ref{Fig13} shows for the example of the \(x=0.26\) and \(x=0.32\) samples how the spectral weight and the scattering rate of the narrow (\(D1\)) and broad (\(D2\)) Drude peaks are affected by the formation of the SDW. 
The scattering rate of the broad Drude-peak remains almost constant and therefore has been fixed to reduce the number of fit parameters. 
Figure~\ref{Fig13}c shows that the scattering rate of the narrow Drude peak is strongly temperature dependent and exhibits a pronounced decrease toward low temperature that is quite similar for both samples. 
The most significant difference between the \(x=0.26\) and \(x=0.32\) samples concerns the spectral weight loss of the Drude peaks in the AF state that occurs due to the SDW formation, see Figs.~\ref{Fig13}(a) and (b). 
At \(x=0.26\), the broad Drude peak shows a pronounced spectral weight loss in the AF state whereas the spectral weight of the narrow Drude-peak remains almost constant or even increases slightly below \(T^N\).
A similar behavior was reported for the undoped parent compound~\cite{DaiPRB2016} and, recently for an underdoped Sr\textsubscript{1-x}Na\textsubscript{x}Fe\textsubscript{2}As\textsubscript{2} crystal that undergoes a corresponding transition from o-AF to i-SCDW order~\cite{YangPRB2019}. 
To the contrary, for the \(x=0.32\) sample in the assigned hedgehod SVC state, the major spectral weight loss involves the narrow Drude-peak (\(D_1\)), whereas the SW of the broad Drude peak (\(D_2\)) remains almost constant. 
Since the narrow Drude peak is believed to arise from the electron-like bands near \(X\), and the broad Drude peak from the hole-like bands near \(\Gamma\), the results in Fig.~\ref{Fig13}a,b suggest that the o-AF and i-SCDW orders are giving rise to gaps primarily on the hole-like bands near \(\Gamma\), whilst the SVC order mostly causes a SDW gap on the electron-like bands near \(X\).

\begin{figure}[htb!]
\includegraphics[width=0.8\columnwidth]{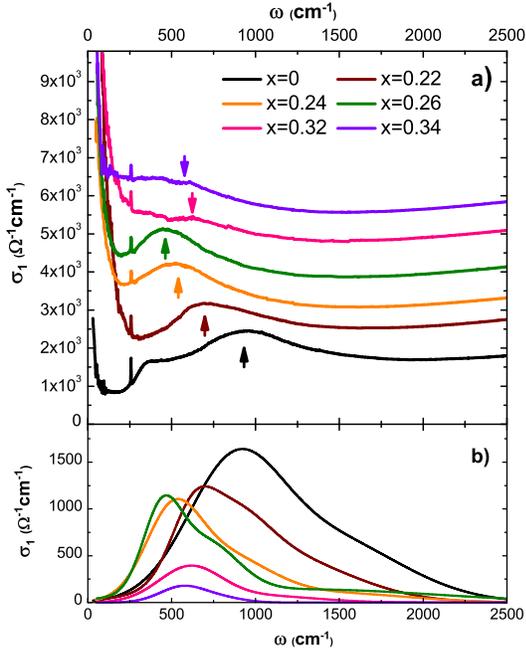}
\caption{(color online) \textbf{(a)} Doping dependence of the optical conductivity in the AF state at 25 K. A vertical offset has been added for clarity. \textbf{(b)} Evolution of the SDW peak as fitted with the Gaussian functions that are described in equation~\ref{eq:3} and shown in Fig.~\ref{Fig11}. The peak intensity decreases continuously as a function of hole doping, whereas the peak frequency exhibits a partial recovery in the suspected SVC state at \(x=0.32\) and \(0.34\).}
\label{Fig14}
\end{figure}

A different trend in the assigned SVC state, as compared to the one in the o-AF and the i-SCDW states, is also evident from the doping dependence of the frequency of the SDW peak. 
Figure~\ref{Fig14}a shows a comparison of the optical conductivity spectra in the AF state at 25K and Fig.~\ref{Fig14}b the evolution of the Gaussian fits of the SDW peak as detailed in Fig.~\ref{Fig11}. 
Whereas the SW of the SDW peak decreases continuously with hole doping (as was already discussed above and shown in Fig.~\ref{Fig11}), the peak frequency also decreases at first in the o-AF and i-SCDW states, from about 700~cm\textsuperscript{-1} at \(x=0.22\) to about 400~cm\textsuperscript{-1} at \(x=0.26\), but then increases again to about 600~cm\textsuperscript{-1} in the SVC state at \(x=0.32\) and \(0.34\). 
Since the SDW peak energy is expected to be proportional to the magnitude of the SDW gap, this anomaly suggests that the average magnitude of the SDW gap in the SVC state exceeds the one in the i-SCDW state. 
The combined evidence from our infrared data thus suggests that the SVC order at \(x=0.32\) and \(0.34\) involves an electron-like band around the \(X\)-point that has quite a large SDW gap but only a weak contribution to the optical spectral weight. 
The latter point can be explained either in terms of a very low concentration or a large effective mass of the charge carriers of this band. 

\begin{figure}[htb!]
\includegraphics[width=1\columnwidth]{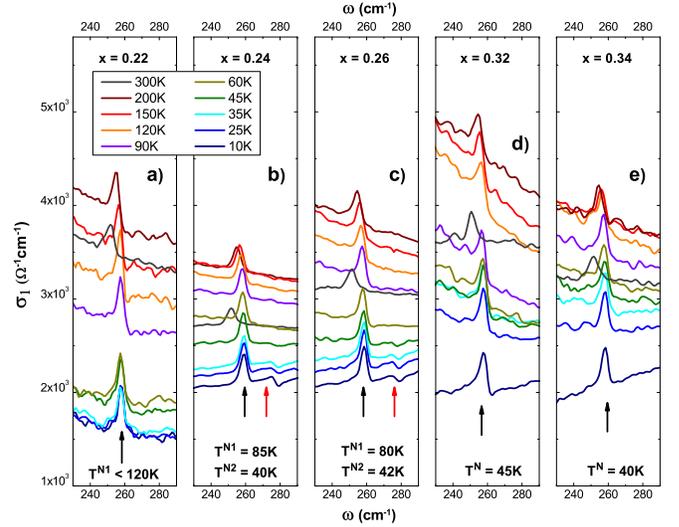}
\caption{(color online) Temperature and doping dependence of the in-plane Fe-As stretching phonon mode near 260~cm\textsuperscript{-1}. A satellite peak is visible here in the i-SCDW state below \(T^{N,2}\approx40K\) at \(x=0.24\) and below \(T^{N,2}\approx42K\) at \(x=0.26\). No sign of such a satellite peak is observed in the other AF states, i.e. in the o-AF state at \(x=0.22\) and the suspected SVC state at \(x=0.32\) and \(0.34\).}
\label{Fig15}
\end{figure}

Additional information about the structural changes in the different AF phases has been obtained from the temperature and doping dependence of the infrared-active phonon mode around 260~cm\textsuperscript{-1} that is summarized in Figure~\ref{Fig15}. 
It was previously reported for BKFA that this in-plane Fe-As stretching mode develops a side band at a slightly higher energy in the i-SCDW state~\cite{MallettPRL2015, MallettPRB2017}.
This new feature was explained in terms of an enlarged unit-cell and a subsequent Brillouin-zone folding due to the presence of two inequivalent Fe sites (with and without a static magnetic moment) in the i-SCDW state \cite{MallettPRL2015}.
Figure~\ref{Fig15} confirms that a corresponding phonon side band occurs in BNFA at \(x=0.24\) and \(0.26\) in terms of an additional peak around 275~cm\textsuperscript{-1} that develops right below \(T^{N,2}\approx40K\). 
Notably, such a satellite peak is not observed in the assigned SVC phase at \(x=0.32\) and \(0.34\). 
This finding confirms that the enlargement of the unit cell and the subsequent BZ folding is unique to the i-SCDW order and emphasizes the distinct nature of the SVC order at \(x=0.32\) and \(0.34\). 
 
\begin{figure*}[htb!]
\includegraphics[width=0.8\textwidth]{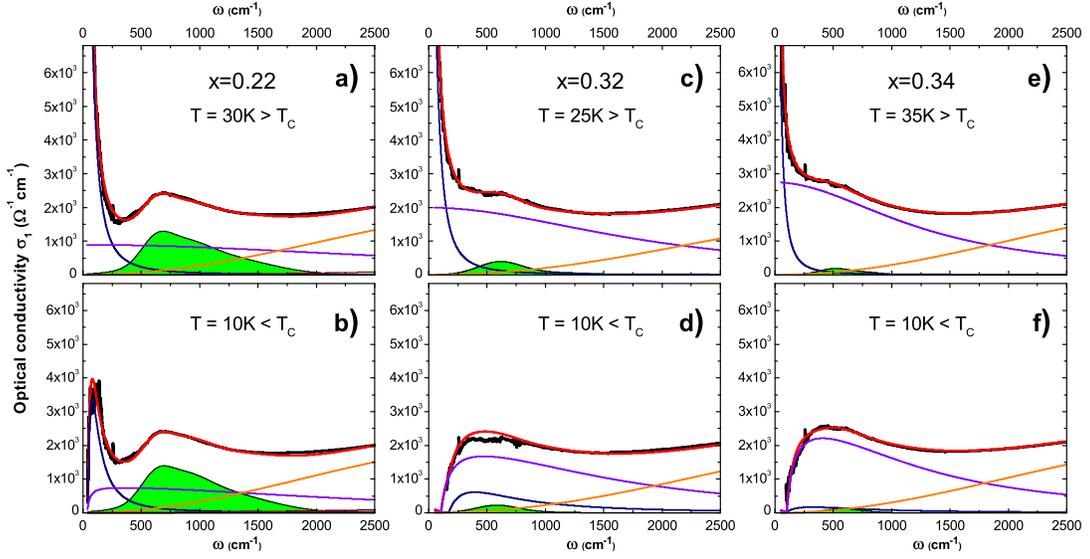}
\caption{(color online) Selected spectra of the optical conductivity and their fitting slightly above and well below \(T_c\) in the o-AF state at \(x=0.22\) (left panels) and in the SVC state at \(x=0.32\) (middle panels) and \(x=0.34\) (right panels). Note that no sign of a SC gap formation has been observed in the corresponding spectra of the samples with \(x=0.24\) and \(0.26\). In the normal state at \(T\approx T_c\), the experimental spectra (black line) have been fitted with two Drude-terms (blue solid lines), a Lorentzian (orange line) and two Gaussian peaks for the SDW pair-breaking peak as discussed in the text. In the SC state at \(T<<T_c\) a Mattis-Bardeen-type isotropic gap function has been added to each Drude-band. The values of the obtained SC gap energies are listed in Table~\ref{Table1}.}
\label{Fig16}
\end{figure*}

Finally, we discuss how the onset of superconductivity affects the spectra of the infrared conductivity in the presence of the different AF orders. 
Figure~\ref{Fig16} shows the corresponding changes to the optical conductivity due to the formation of the SC gap(s) for the samples in the o-AF state at \(x=0.22\) and in the assigned SVC state at \(x=0.32\) and \(0.34\). 
For the samples in the i-SCDW state at \(x=0.24\) and \(0.26\) no sign of the formation of a SC gap and a related delta-function at zero frequency due to a SC condensate has been observed down to the lowest measured temperature of 10K (see Figs.~\ref{Fig10}c-f). 
As shown in Fig.~\ref{Fig5}, this is despite of a bulk SC transition at \(T_c\approx 12K\) at \(x=0.24\) measured with the specific heat.
Our infrared data thus reflect a strong suppression of the superconducting response due to the competition with the i-SCDW order that is more severe than in the o-AF and the assigned SVC states. 
Finally, note that clear signatures of a SC energy gap have very recently been reported for a similar BNFA sample for which \(T_c\) was somewhat higher and the measurements were performed to a lower temperature of 5K~\cite{YangPRB2019}.

The upper panels of Fig.~\ref{Fig16} show the spectra in the normal state slightly above \(T_c\) for which the fitting has already been shown in Fig.~\ref{Fig11}. 
The lower panels display the corresponding spectra and their fitting in the SC state. 
Here the optical conductivity at low frequency (below 50~cm\textsuperscript{-1} at \(x=0.22\), 120~cm\textsuperscript{-1} at \(x=0.32\) and 100~cm\textsuperscript{-1} at \(x=0.34\)) is strongly suppressed due to the opening of the superconducting energy gap(s). 
This SC gap formation has been accounted for using a Mattis-Bardeen-type model that allows for isotropic gaps of different magnitude on the narrow and the broad Drude-bands.   

For the \(x=0.22\) crystal, for which the SC state coexists with a strong o-AF order, there are clear signs of the SC gap formation below \(T_c\approx15K\), i.e. Figure~\ref{Fig16}b reveals a strong suppression of the optical conductivity toward low frequency. 
The SC gap edge is also evident from the bare reflectivity spectrum in the inset of Fig.~\ref{Fig10}(a). 
The obtained gap energies amount to \(2\Delta^{SC}\approx4.4\) meV and 5.2 meV for the broad and narrow Drude-bands, respectively, and ratios of \(2\Delta^{SC}/k_B T_c\approx2.87\) and 3.35 that compare rather well with the prediction of the weak coupling BCS theory of \(2\Delta^{SC}/k_B T_c=3.54\). 

A strong increase of the SC gap energy is observed for the samples in the SVC state at \(x=0.32\) (Fig.~\ref{Fig16}, middle panel) and \(x=0.34\) (Fig.~\ref{Fig16}, right panel) for which the overall shape of the SC spectra is quite similar to the one of optimally doped BKFA~\cite{WuPRB2009, MallettPRB2017, BingXuPRB2016, CharnukhaNComm2011, CharnukhaPRB2011}.
Here \(2\Delta^{SC}\) for the narrow and broad Drude-bands amounts to 19.8 meV and 13 meV at \(x=0.32\), and 30 meV and 12 meV at \(x=0.34\), respectively (see also Table~\ref{Table1}). 
Similar to BCFA and BKFA~\cite{KimPRB2010, HeumenEPL2010, BingXuPRB2017, YangPRB2019} and also BCKFA~\cite{YangPRB2017}, the larger SC gap is assigned to the narrow Drude-peak, which supposedly originates from the electron-like bands near the \(X\)-point of the Brillouin zone. 

Finally, we derived the SC plasma frequency, \(\Omega^2_{pS}\), and the related ratio of the condensate density to the effective band mass, \(\frac{n_s}{m^{*}_{ab}}\), from the analysis of the missing spectral weight using the Ferrell-Glover-Tinkham (FGT) sum rule:
\begin{equation}
	\label{eq4}\Omega^2_{pS}=\frac{Z_0}{\pi^2}\int_{0^+}^{\omega_c}\left[\sigma_1(\omega,T\sim T_c)-\sigma_1(\omega,T<<T_c)\right]d\omega,
\end{equation} 
where the upper cutoff frequency \(\omega_c\) has been chosen such that the optical conductivity in the normal and SC states is almost identical above \(\omega_c\). 

Alternatively, the superfluid density has been determined from the analysis of the inductive response in the imaginary part of the optical conductivity, \(\sigma_2\), according to:
\begin{equation}
	\label{eq5}\frac{n_s}{m^{*}_{ab}}=\Omega^2_{pS}=\frac{Z_0}{2\pi}\omega\sigma_{2S}(\omega).
\end{equation} 	
Here the contribution of the regular response to \(\sigma_2\), due to the excitation of unpaired carriers, has been subtracted as described in Refs.~\cite{DordevicPRB2002, ZimmersPRB2004}. 
Both methods yield consistent values of the SC condensate density, \(\frac{n_s}{m^{*}_{ab}}\), and the related magnetic penetration depth, \(\lambda\), that are listed in Table~\ref{Table1}.

\begin{table*}
\caption{Values of the SC gaps of the narrow and broad Drude bands and of the SC plasma frequency and magnetic penetration depth as obtained from the optical data in the SC state for the samples with \(x=0.22\), \(0.32\), and \(0.34\).}
\begin{ruledtabular}
\begin{tabular}{ ccccccccc}
x (Na) & \(2\Delta^{SC}_{narrow}\) (meV) & \(2\Delta^{SC}_{narrow}/k_B T\) & \(2\Delta^{SC}_{broad}\) (meV) & \(2\Delta^{SC}_{broad}/k_B T\) & \(\Omega^2_{pS}\) (cm$^{-2}$)& $\lambda$ (nm) \\
0.22 & 5.21 & 3.35 & 4.46 & 2.87 & 4.9 \(\cdot 10^7\) & 227 \\
0.32 & 19.84 & 10.46 & 13.14 & 6.93 & 6.2 \(\cdot 10^7\) & 202 \\
0.34 & 30.75 & 14.27 & 12.65 & 5.87 & 5.4 \(\cdot 10^7\) & 216 \\
\end{tabular}
\end{ruledtabular}
\label{Table1}
\end{table*}

\section{Discussion and Summary}

We have performed a combined \(\mu\)SR- and infrared spectroscopy study of the magnetic part of the phase diagram of the hole doped BNFA system. 
We have confirmed that the so-called double-\textbf{Q} AF state with an inhomogeneous spin-charge density wave (i-SCDW) order exists in a sizable doping range where it persists to the lowest measured temperature, i.e. even below \(T_c\). 
This is different from BKFA where the i-SCDW order exists only in a rather narrow doping regime and exhibits a re-entrance to an o-AF state at low temperature~\cite{BohmerNComm2015, MallettPRL2015, MallettEPL2015, MallettPRB2017}.
Otherwise, we observed the same signatures of the i-SCDW state as in BKFA. 
This concerns the reorientation of the spins from an in-plane direction in the o-AF state to an out-of-plane one in the i-SCDW state. 
We also observed a satellite peak of the infrared-active Fe-As stretching phonon mode, which signals a folding of the Brilluoin zone due to an enlarged unit cell in the i-SCDW state. 
In the infrared spectra, at the lowest measured temperature of 10K, no sign of a bulk-like SC response has been seen with infrared spectroscopy in the i-SCDW state of the BNFA crystals with \(x=0.24\) and 0.26, for which a bulk SC transition is evident from specific heat.
This suggests that the superconducting response is strongly suppressed by the competition with i-SCDW order.

We also obtained evidence for a new type of t-AF state that is likely a hedgehog-type spin-vortex-crystal (SVC) order. 
This new AF phase shows up at a higher hole doping level than the i-SCDW phase and persists until the static magnetism is fully suppressed at optimum doping. 
This additional magnetic phase in the BNFA phase diagram was first discovered with thermal expansion measurements where it shows up in terms of a very small orthorhombic distortion~\cite{WangPRB2016}. 
Accordingly, it has been interpreted in terms of an o-AF order that is either very weak, strongly incommensurate, or inhomogeneous. 
To the contrary, our \(\mu\)SR data establish that this AF state is bulk-like, more or less commensurate and has a surprisingly large magnetic moment (at least at \(T>T_c\)). 
Due to its almost tetragonal structure and since the \(\mu\)SR data reveal magnetic moments that are rather large and oriented along the FeAs planes, we have assigned this new AF order to an orthomagnetic double-\textbf{Q} state, in particular, to the hedgehog-type spin vortex crystal (SVC) structure. 
This SVC state was previously only observed in the K,Ca-1144 structure where it is believed to be stabilized by the reduced disorder and/or the breaking of the glide-plane symmetry of the FeAs layers due to the alternating layers of Ca and K ions~\cite{MeierQMat2018, DingPRL2018}.
It is therefore interesting that this kind of SVC order also occurs in the present BNFA system for which the Na and Ba ions are randomly distributed. 
Another remarkable feature of this SVC order is its very strong competition with superconductivity which leads to a large reduction of the magnetic moment (at \(x=0.32\)) and even of the magnetic volume fraction (at \(x=0.34\)). 
A similarly large suppression of magnetic order due to the onset of SC was so far only observed in BFCA crystals close to optimum doping for which a very weak incommensurate o-AF order exhibits a re-entrance into a non-magnetic state below \(T_c\)~\cite{PrattPRL2011}.

Another interesting aspect of our present work emerges from the comparison of the doping evolution of the magnitude of the ordered magnetic moment as deduced from the local magnetic field in the \(\mu\)SR experiment and the SW of the SDW peak in the infrared spectroscopy data. 
The trends can be seen in Figure~\ref{Fig7} which compares the doping evolution of the AF moment, normalized to the one of the undoped parent compound, as seen with \(\mu\)SR (which probes the total ordered magnetic moment) and infrared spectroscopy (which is only sensitive to the itinerant moment). 
The solid blue symbols show the value of the AF order parameter as obtained from the local magnetic field at the muon site. 
The solid orange symbols show the corresponding values of the SW of the SDW peak. 
In both cases, the amplitude of the magnetic moment decreases continuously with doping, but the decrease is considerably stronger for the itinerant moments deduced from the infrared data, than for the total magnetic moment seen with \(\mu\)SR. 
This might indicate that the ordered magnetic order has a mixed character with contribution from itinerant and from localized moments. 
The different trends of the optics and \(\mu\)SR data thus could be explained if the magnetic moments are more strongly localized as the hole doping increases. 
An alternative, and to our opinion more likely explanation is in terms of a change of the effective mass of the itinerant charge carriers that are gapped by the SDW. 
The very small SW of the SDW peak in the SVC state, as compared to the large local magnetic field in the \(\mu\)SR experiment, thus implies that the SDW gap develops on a flat band with a rather large effective mass. 
Note that such a scenario, that the SDW develops on different parts of the Fermi-surface in the SVC state, as compared to the o-AF and i-SCDW states, is consistent with the data in Figure~\ref{Fig13} which show that the SDW peak obtains a major part of its SW from the narrow Drude peak, rather than from the broad one, as in the o-AF and i-SCDW states. 
This scenario could be probed e.g. by future ARPES studies on such BNFA crystals in the i-SCDW and SVC states.

\appendix

\section{Muon site calculation}

The space group symmetry of Ba\textsubscript{1-x}K\textsubscript{x}Fe\textsubscript{2}As\textsubscript{2} (BKFA) and Ba\textsubscript{1-x}Na\textsubscript{x}Fe\textsubscript{2}As\textsubscript{2} (BNFA) in the paramagnetic phase is I4/mmm with one formula unit (\(Z=1\)) in the primitive cell. 
The Ba ions reside in the 1a – position \((0,0,0)\), As in the 2e – position \((0,0,zAs)\) and Fe in the 2d – position \((0,1/2,1/4)\).
Note that the crystallographic unit cell differs from the primitive cell which is built by primitive translations: \(a_1=(-a/2;b/2;c/2)=(-\tau;\tau;\tau_c)\), \(a_2=(a/2;-b/2;c/2)=(\tau;-\tau;\tau_c)\), \(a_3=(a/2;b/2,-c/2) =(\tau;\tau;-\tau_c)\).
In the following we analyze the position of the muon stopping sites for a K content of \(x=0.2465\) with the structural room temperature data: \(a=b=3.9343\) \AA, \(c=13.2061\) \AA, and \(z_{As} = 0.35408\) \AA.
We assume here that these muon sites do not strongly change when the K-content is varied or when K is replaced by Na.

We used a modified Thomas Fermi approach~\cite{ReznikPRB1995} that allows a direct determination of the self-consistent distribution of the valence electron density from which the electrostatic potential can be restored. 
The local, interstitial minima of this electrostatic potential are identified as muon stopping sites. 

For the same purposes we performed more elaborated \textit{ab initio} calculations within the framework of density functional theory (DFT). 
We applied the all-electron full-potential linearized augmented plane wave method (Elk code)~\cite{Elk} with the local spin density approximation~\cite{PerdewPRB1992} for the exchange correlation potential and with the revised generalized gradient approximation of Perdew-Burke-Ernzerhof~\cite{PerdewPRL2008}.
The calculations were performed on a \(9\times 9\times 6\) grid which corresponds to 60 points in the irreducible Brillouin zone. 
In both approaches we used a supercell \(2a\times 2b\times c\) and supposed \(x=0.25\) (e.g. Ba\textsubscript{0.75}K\textsubscript{0.25}Fe\textsubscript{2}As\textsubscript{2}).
This allows one to explicitly incorporate K ions which were positioned in the supercell at coordinates K(1) - \((a, b, 0)\) and K(2) - \((3/2a, 3/2b, 1/2c)\).

The DFT and modified Thomas Fermi approaches give almost the same answers. 
We observed three possible types of muon sites. 
Two of them are located on the line along the \(c\)-direction connecting the nearest Ba or K and As ions at the coordinates \((0, 0, z_\mu)\) with \(z_\mu =0.191\) for Ba and \(z_\mu =0.170\) for K.  
In the I4/mmm setting these muon sites have a 2e – local point symmetry (4mm), i.e. the same as the As ions.
We have verified that the dipolar fields from a given magnetic structure of the Fe moments have nearly the same magnitudes at these two positions.
Accordingly, in the dipolar field calculations we discuss only one type of muon stopping site. 

The third muon site is located in the Ba \(ab\)-plane close to the line connecting the As-As ions along the \(c\)-direction.
In the I4/mmm setting it has a rather high 4j – local point symmetry (m2m) at the coordinates \((0.4,0.5,0)\). 
Its electrostatic potential is roughly 20\% less than the potential of the previous two sites.
Accordingly, this site should be less populated in the \(\mu\)SR experiment. 
The probability of the occupation of this secondary site, as compared to the one of the majority site, we calculate to be 0.24 which agrees rather well with the experimental amplitude ratio of \(A^{os}_{2}/A^{os}_{1}\approx 0.2\) (see Fig.~2 of the Ref.~\cite{MallettEPL2015}).
The qualitative changes of the local dipole fields on this minority site at the o-AF to t-AF transition are very similar to the ones on the majority muon sites. 
Accordingly, in the following and in the paper we do not further discuss this minority muon site and focus instead on the changes of the local dipole field on the majority muon site. 

\begin{figure}[htb!]
\includegraphics[width=0.8\columnwidth]{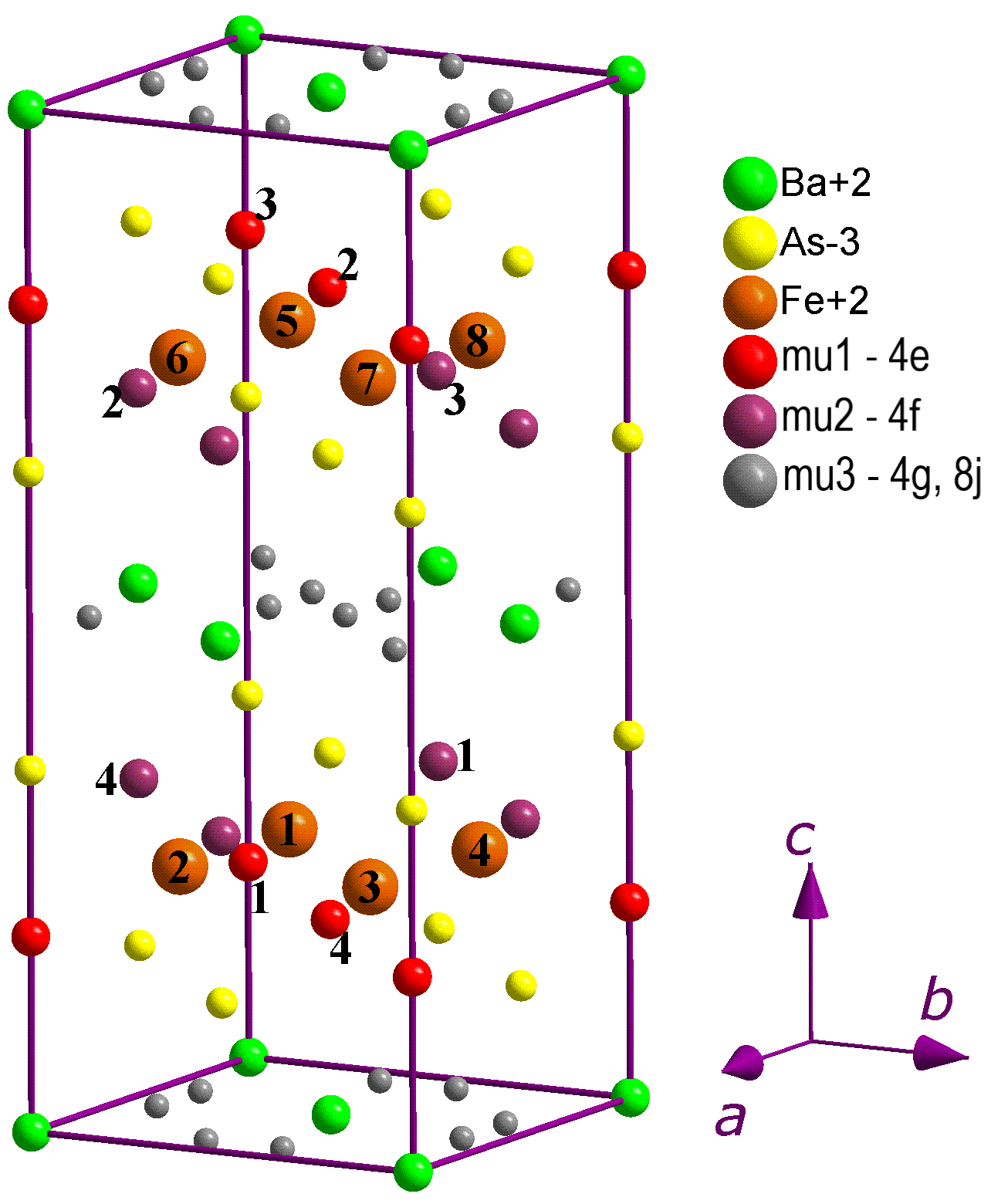}
\caption{(color online) Sketch of the unit cell of Ba\textsubscript{1-x}K\textsubscript{x}Fe\textsubscript{2}As\textsubscript{2} in the tetragonal subgroup P4/mbm of the space group I4/mmm. Atoms and muon stopping sites are in the positions; Ba\textsubscript{1} - 2a \((0,0,0)\), Ba\textsubscript{2} – 2c \((0,1/2,1/2)\), As\textsubscript{1}/\(\mu_1\) – 4e \((0,0,z_1(As/\mu)\) with \(z_1(As)=0.35408\) and \(z_1(\mu)=0.188\),  As\textsubscript{2}/\(\mu_2\) – 4f \((0,1/2, z_2(As/\mu))\) with \(z_2(As)=0.14592\) and \(z_2(\mu)=0.312\),  Fe – 8k \((x,x+1/2,z)\) with \(x=1/4\), \(z=1/4\),  \(\mu_3\) - 4g \((0.4,0.9,0)\) and 4g – \((0.1,0.6,0)\) and 8j – \((0.1,0.1,0.5)\). The enumeration of the Fe- and \(\mu_1\) and \(\mu_2\) sites is indicated. }
\label{Fig17}
\end{figure}

\section{Calculation of the dipolar field at the muon site}

To unify the description of the possible magnetic structures in the tetragonal phase of Ba\textsubscript{1-x}K\textsubscript{x}Fe\textsubscript{2}As\textsubscript{2} (BKFA) and Ba\textsubscript{1-x}Na\textsubscript{x}Fe\textsubscript{2}As\textsubscript{2} (BNFA) we used the space group P4/mbm N127 that is the subgroup of index 4 of the parent group I4/mmm N139. 
This choice is dictated by the expected four-fold increase of the magnetic unit cell as compared to the parent I4/mmm primitive cell that is caused by the lowering of the translation symmetry.
The P4/mbm subgroup has the same origin as the parent group I4/mmm and the basis \((\textbf{a},\textbf{b},\textbf{c})\) that is rotated by 45\(\textsuperscript{o}\) in the ab-plane as compared to the I4/mmm basis \((\textbf{a}\prime,\textbf{b}\prime,\textbf{c}\prime)\) with \(a=b=2a'\) and \(c=c'\). 
In the P4/mbm setting the eight Fe atoms in the unit cell are in the – 8k \((x,x+1/2,z)\) position.
The 2e position of the As atoms and of the muon site in the I4/mmm notation are divided in the P4/mbm setting into the 4e – \((0,0,z_1(As/\mu))\) and 4f - \((0,1/2, z_2(As/\mu))\) positions. 
Respectively, the 4j position of the third muon site in the I4/mmm notation are divided in the P4/mbm setting into the 8j – \((x,y,1/2)\) and two 4g - \((x,x+1/2,0)\) positions.
The primitive cell of BKFA in the P4/mbm setting is shown in Fig.~\ref{Fig17}.

The symmetry consideration of the possible \textbf{2k-} and \textbf{1k-} (or double-\textbf{Q} and single-\textbf{Q}) magnetic structures is based on the so called representation analysis of the magnetic degrees of freedom that are real and located on the magnetic ions and that are virtually assigned on the muon stopping sites~\cite{Bertaut1971,Opechowski1971,Izyumov1991}.
The magnetic degrees of freedom, for a set of atoms at a given Wyckoff position, form a magnetic representation which is reducible and can be decomposed into irreducible representations (IR). 
The possible magnetic structures can be presented in terms of a linear combination of magnetic moments, \textbf{\textit{L}}, which transform under the symmetry operations as basic functions of a given IR. 
This is in accordance with the Landau concept that only one IR is realized at a phase transition for which \textbf{\textit{L}} is a nonzero order parameter in the low symmetry phase.

Purely based on symmetry arguments one can make the following strict predictions for the local magnetic field that is seen in a zero-field \(\mu\)SR experiment. 
The complex magnetic structure does not give rise to a finite magnetic field at the muon site if the IR of its order parameter does not enter into the decomposition of the magnetic representation for the muon site. 

This circumstance is illustrated below for the possible magnetic structures in the tetragonal phase of BKFA. 
For the following analysis it is important to note that the lowering of the translation symmetry in the \textbf{2k}-structures is already accounted for by using a four times enlarged primitive unit cell. 
In the P4/mbm setting thus we can perform the symmetry treatment for the Fe- and muon-site magnetic representations for the propagation vector \(K_0=(0,0,0)\). 

To represent the order parameters of the respective \textbf{2k}-magnetic structures, which can arise in the I4/mmm setting with the propagation vectors \(k_1=(1/2,1/2,0)\) and \(k_2=(-1/2,1/2,0)\), we introduce the following linear combinations \textbf{\textit{L}} of the magnetic iron moments in the P4/mbm setting with \(K_0=(0,0,0)\).
\begin{equation}
\begin{aligned}
&\begin{array}{c}
\vec{F}^{(\pm)}=1 / 8 [\left(\vec{m}_{1}+\vec{m}_{2}+\vec{m}_{3}+\vec{m}_{4}\right) \\
\pm\left(\vec{m}_{5}+\vec{m}_{6}+\vec{m}_{7}+\vec{m}_{8}\right) ]; \\
\vec{L}_{1}^{(\pm)}=1 / 8 [\left(\vec{m}_{1}+\vec{m}_{2}-\vec{m}_{3}-\vec{m}_{4}\right) \\
\pm\left(\vec{m}_{5}+\vec{m}_{6}-\vec{m}_{7}-\vec{m}_{8}\right)];\\
\vec{L}_{2}^{(\pm)}=1 / 8 [\left(\vec{m}_{1}-\vec{m}_{2}+\vec{m}_{3}-\vec{m}_{4}\right) \\
\pm\left(\vec{m}_{5}-\vec{m}_{6}+\vec{m}_{7}-\vec{m}_{8}\right)];\\
\vec{L}_{3}^{(\pm)}=1 / 8 [\left(\vec{m}_{1}-\vec{m}_{2}-\vec{m}_{3}+\vec{m}_{4}\right) \\
 \pm\left(\vec{m}_{5}-\vec{m}_{6}-\vec{m}_{7}+\vec{m}_{8}\right)].
\end{array}
\end{aligned}
\label{eqAppB1}
\end{equation}
The magnetic order parameters \textbf{\textit{L}} consist of the Fourier components of the magnetic propagation vector, \textbf{\textit{K\textsubscript{0}}}, in terms of the sub-lattice magnetic moments \(m_\alpha\) with \(\alpha=1\div 8\). 
Similarly one can introduce linear combinations of the \(K_0\) - Fourier components of the magnetic fields \(B^{I,II}_{\alpha} (\alpha=1\div 4)\) at the muon positions with 4e and 4f site symmetry that are enumerated by I and II respectively.
The respective staggered magnetic fields at these muon sites have the form:
\begin{equation}
\begin{aligned}
&\vec{F}^{(I,II)}=\frac{1}{4}\left(\vec{B}_{1}^{(I,II)}+\vec{B}_{2}^{(I,II)}+\vec{B}_{3}^{(I,II)}+\vec{B}_{4}^{(I,II)}\right);\\
&L_{1}^{(I,II)}=\frac{1}{4}\left(\vec{B}_{1}^{(I,II)}+\vec{B}_{2}^{(I,II)}-\vec{B}_{3}^{(I,II)}-\vec{B}_{4}^{(I,II)}\right);\\
&L_{2}^{(I,II)}=\frac{1}{4}\left(\vec{B}_{1}^{(I,II)}-\vec{B}_{2}^{(I,II)}+\vec{B}_{3}^{(I,II)}-\vec{B}_{4}^{(I,II)}\right);\\
&L_{3}^{(I,II)}=\frac{1}{4}\left(\vec{B}_{1}^{(I,II)}-\vec{B}_{2}^{(I,II)}-\vec{B}_{3}^{(I,II)}+\vec{B}_{4}^{(I,II)}\right).
\end{aligned}
\label{eqAppB2}
\end{equation}
The quantities defined in Eqs.~\ref{eqAppB1} and \ref{eqAppB2} can serve as the basic functions of the irreducible representations of the P4/mbm group with propagation vector \(K_0=(0,0,0)\). 
The attribution of these basic functions to the IR of the tetragonal group P4/mbm is as shown in Table~\ref{T1AppB}.
The possible eight noncollinear spin vortex crystal structures which are allowed by the double-\textbf{Q} magnetic order are described by the \(\tau_{1}-\tau_{8}\) irreducible repesentations.

The following examples illustrate how to read the data of Table~\ref{T1AppB}.
The magnetic structures which can be realized with the iron order parameters of a given IR give rise to staggered fields at the muon sites that transform by the same IR.
For example, the magnetic structure which transforms according to the IR \(\tau_5-B_{1g}\) consists of the two order parameters \(L^{(-)}_{3x}-L^{(-)}_{1x}\) and \(L^{(+)}_{2z}\). 
According to Table~\ref{T1AppB}, both order parameters \(L^{(-)}_{3x}-L^{(-)}_{1x}\) and \(L^{(+)}_{2z}\) do not create finite dipolar fields at the 4e muon stopping sites.
At the same time, at the 4f muon stopping sites they both create dipolar fields that are directed along the c-axis and have the same staggered structure \(L^{II}_{2z}\).
This is a strict result if we take the iron coordinates in the form Fe – 8k \((x,x+1/2,z)\).
However, there is the starting symmetry I4/mmm which we can reproduce by taking the iron coordinates as \(x=1/4\), \(z=1/4\) so that we get 8k \((1/4,3/4,1/4)\). 
This additional, internal symmetry leads to the disappearance of the magnetic fields at some of the muon stopping sites.  
\begin{table*}
\caption{Symmetry of the order parameters of the possible Fe-based magnetic phases and the symmetry and magnitude of the respective staggered magnetic fields from Eq.~\ref{eqAppB2} at the muon sites in the tetragonal phase of Ba\textsubscript{1-x}K\textsubscript{x}Fe\textsubscript{2}As\textsubscript{2} in P4/mbm setting for the magnetic propagation vector \(K_0=(0,0,0)\)}
\begin{ruledtabular}
\begin{tabular}{m{3.5em} m{9em} m{5em} m{5em} m{5em} m{5em}}
P4/mbm & \multicolumn{5}{c}{\(K_0=(0;0;0)\)} \\
\hline
IR & \textbf{Fe} order parameters & Fields at \newline muon sites \newline 4e I & Frequency at 1~\(\mu_B/Fe\) \newline in  MHz & Fields at \newline muon sites \newline 4f II & Frequency at 1~\(\mu_B/Fe\) \newline in  MHz \\
\hline
\(\tau_{1}-A_{1 g}\) & \(L_{3 x}^{(-)}+L_{1 y}^{(-)}\) & \(L_{2 z}^{(I)}\) & 52.84 & - & 0 \\
\(\tau_{2}-A_{1 u}\) & \colorbox{yellow}{\(L_{1 x}^{(+)}-L_{3 y}^{(+)}\)}; \(F_{z}^{(-)}\) & \(L_{3 z}^{(I)}\) & \colorbox{yellow}{0} & \(L_{3 z}^{(II)}\) & \colorbox{yellow}{0} \\
\(\tau_{3}-A_{2 g}\) & \colorbox{yellow}{\(L_{1 x}^{(-)}-L_{3 y}^{(-)}\)}; \(F_{z}^{(+)}\) & \(F_{z}^{(I)}\) & \colorbox{yellow}{0} & \(F_{z}^{(II)}\) & \colorbox{yellow}{0} \\
\(\tau_{4}-A_{2 u}\) & \(L_{3 x}^{(+)}+L_{1 y}^{(+)}\) & \(L_{1 z}^{(I)}\) & 52.03 & - & 0 \\
\(\tau_{5}-B_{1 g}\) & \(L_{3 x}^{(-)}-L_{1 y}^{(-)}\); \colorbox{yellow}{\(L_{2 z}^{(+)}\)} & - & 0 & \(L_{2 z}^{(II)}\) & 52.84 \\
\(\tau_{6}-B_{1 u}\) & \colorbox{magenta}{\(L_{1 x}^{(+)}+L_{3 y}^{(+)}\)} & - & 0 & - & 0 \\
\(\tau_{7}-B_{2 g}\) & \colorbox{magenta}{\(L_{1 x}^{(-)}+L_{3 y}^{(-)}\)} & - & 0 & - & 0 \\
\(\tau_{8}-B_{2 u}\) & \(L_{3 x}^{(+)}-L_{1 y}^{(+)}\); \colorbox{yellow}{\(L_{2 z}^{(-)}\)} & - & 0 & \(L_{1 z}^{(II)}\) & 52.03 \\
\(\tau_{9}-E_{g}\) & \(\left\{\begin{array}{c}F_{x}^{(+)} \\ -F_{y}^{(+)}\end{array}\right.\); \(\left\{\begin{array}{c}L_{2 y}^{(+)} \\ -L_{2 x}^{(+)}\end{array}\right.\); \(\left\{\begin{array}{l}L_{1 z}^{(-)} \\ L_{3 z}^{(-)}\end{array}\right.\) & \(\left\{\begin{array}{c}F_{x}^{(I)} \\ -F_{y}^{(I)}\end{array}\right.\); \(\left\{\begin{array}{l}L_{2 y}^{(I)} \\ L_{2 x}^{(I)}\end{array}\right.\) & & \(\left\{\begin{array}{c}F_{x}^{(II)} \\ -F_{y}^{(II)}\end{array}\right.\); \(\left\{\begin{array}{l}L_{2y}^{(II)} \\ -L_{2x}^{(II)}\end{array}\right.\) & \\
\(\tau_{10}-E_{u}\) & \(\left\{\begin{array}{c}L_{2x}^{(-)} \\ L_{2y}^{(-)}\end{array}\right.\); \(\left\{\begin{array}{c}F_{y}^{(-)} \\ F_{x}^{(-)}\end{array}\right.\); \(\left\{\begin{array}{l}L_{3z}^{(+)} \\ -L_{1z}^{(+)}\end{array}\right.\) & \(\left\{\begin{array}{c}L_{1x}^{(I)} \\ -L_{1y}^{(I)}\end{array}\right.\); \(\left\{\begin{array}{l}L_{3y}^{(I)} \\ L_{3x}^{(I)}\end{array}\right.\) & & \(\left\{\begin{array}{c}L_{1x}^{(II)} \\ L_{1y}^{(II)}\end{array}\right.\); \(\left\{\begin{array}{l}L_{3y}^{(II)} \\ L_{3x}^{(II)}\end{array}\right.\) & \\
\end{tabular}
\end{ruledtabular}
\label{T1AppB}
\end{table*}

\begin{figure}[htb!]
\includegraphics[width=0.95\columnwidth]{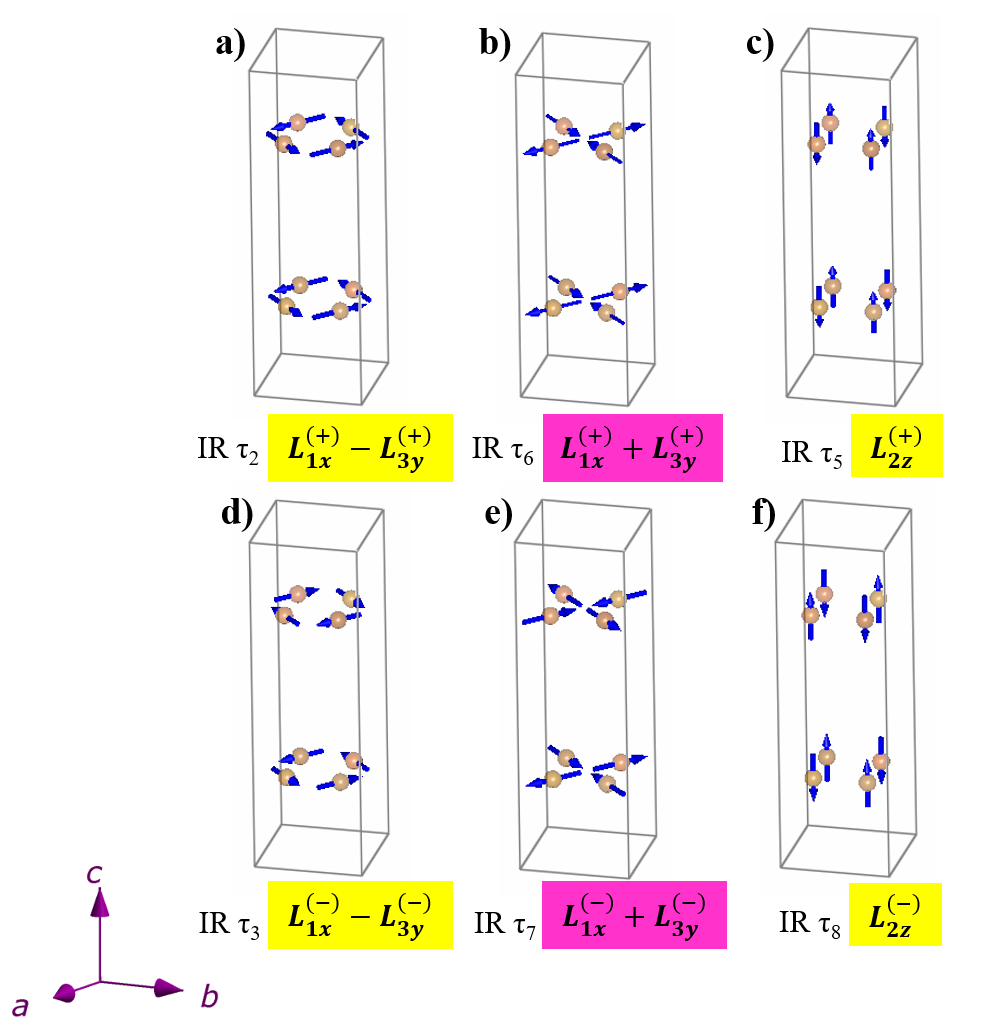}
\caption{(color online) SVC loop double-\textbf{Q} magnetic structures in P4/mbm setting which preserve the \(C4\) symmetry and do not create a magnetic dipole field at the muon sites. Only the iron atoms are shown. The structures in \textbf{a)}, \textbf{b)} and \textbf{c)} exhibit a FM order along the \(c\)-axis, the ones in \textbf{d)}, \textbf{e)}, \textbf{f)} a corresponding AFM order. }
\label{Fig18}
\end{figure}

The magnetic structures (order parameters), which do not give rise to a finite magnetic field at the muon site for \(x=1/4\), \(z=1/4\), are marked in "yellow".
The "pink" color denotes the magnetic structures (order parameters) that cannot be detected by \(\mu\)SR for the given 4e and 4f muon stopping sites, even for an arbitrary choice of the \(x\)-and \(z\)-coordinates in Fe – 8k \((x,x+1/2,z)\). 
All of these structures are illustrated in Fig.~\ref{Fig18}.

Note that structures marked in “yellow” can give rise to small, finite fields at the muon sites in the case of small, static deviations of the iron coordinates from the values \(x=1/4\) and \(z=1/4\). 
In this case the local fields and the resulting \(\mu\)SR precession frequencies will be more or less proportional to the amplitude of the deviations. 

Below we summarize the outcome of the dipole field calculations for the magnetic order parameters with AF order for the case Fe – 8k \((1/4,3/4,1/4)\).
The magnetic fields are given in units of MHz, corresponding to the \(\mu\)SR precession frequency, \(\nu_\mu=\frac{\gamma_\mu}{2\pi}\cdot B_\mu\), and the magnetic order parameters (linear combinations from Eq.~\ref{eqAppB1}) in units of \(\mu_B\).

The fields at the 4e-muon sites with coordinates \((0.0,0.0,0.1880)\) are:
\begin{equation}
\begin{aligned}
&\left(\begin{array}{c}
B_{x} \\
B_{y} \\
B_{z}
\end{array}\right)=\left(\begin{array}{ccc}
28.52 & 0 & 0 \\
0 & 28.52 & 0 \\
0 & 0 & -57.04
\end{array}\right)\left(\begin{array}{l}
F_{x}^{(-)} \\
F_{y}^{(-)} \\
F_{z}^{(-)}
\end{array}\right)\\
&+\left(\begin{array}{ccc}
0 & 0 & 0 \\
0 & 0 & 37.36 \\
0 & 37.36 & 0
\end{array}\right)\left(\begin{array}{l}
L_{1 x}^{(-)} \\
L_{1 y}^{(-)} \\
L_{12}^{(-)}
\end{array}\right)+\left(\begin{array}{ccc}
0 & 59.96 & 0 \\
59.96 & 0 & 0 \\
0 & 0 & 0
\end{array}\right)\\
&\times\left(\begin{array}{c}
L_{2 x}^{(-)} \\
L_{2 y}^{(-)} \\
L_{2 z}^{(-)}
\end{array}\right)+\left(\begin{array}{ccc}
0 & 0 & 37.36 \\
0 & 0 & 0 \\
37.36 & 0 & 0
\end{array}\right)\left(\begin{array}{l}
L_{3 x}^{(-)} \\
L_{3 y}^{(-)} \\
L_{3 z}^{(-)}
\end{array}\right).
\end{aligned}
\label{eqAppB3}
\end{equation}

The fields at the 4f-muon sites with coordinates \((0.5,0.0,0.312)\) are:
\begin{equation}
\begin{aligned}
&\left(\begin{array}{c}
B_{x} \\
B_{y} \\
B_{z}
\end{array}\right)=\left(\begin{array}{ccc}
-28.52 & 0 & 0 \\
0 & -28.52 & 0 \\
0 & 0 & 57.04
\end{array}\right)\left(\begin{array}{l}
F_{x}^{(-)} \\
F_{y}^{(-)} \\
F_{z}^{(-)}
\end{array}\right)+\\
&\left(\begin{array}{ccc}
0 & 0 & 0 \\
0 & 0 & -37.36 \\
0 & -37.36 & 0
\end{array}\right)\left(\begin{array}{c}
L_{1 x}^{(-)} \\
L_{1 y}^{(-)} \\
L_{1 z}^{(-)}
\end{array}\right)+\left(\begin{array}{ccc}
0 & 59.96 & 0 \\
59.96 & 0 & 0 \\
0 & 0 & 0
\end{array}\right)\\
&\times\left(\begin{array}{c}
L_{2 x}^{(-)} \\
L_{2 y}^{(-)} \\
L_{2 z}^{(-)}
\end{array}\right)+\left(\begin{array}{ccc}
0 & 0 & 37.36 \\
0 & 0 & 0 \\
37.36 & 0 & 0
\end{array}\right)\left(\begin{array}{c}
L_{3 x}^{(-)} \\
L_{3 y}^{(-)} \\
L_{3 z}^{(-)}
\end{array}\right).
\end{aligned}
\label{eqAppB4}
\end{equation}

In the following Figure~\ref{Fig18} we show the magnetic structures which do not give rise to a magnetic field at the muon site and thus to a finite \(\mu\)SR precession frequency. 
These structures are therefore not compatible with our experimental data in the t-AF state. 
Interestingly, all of them belong to so called loop-type SVC structures.

\begin{figure}[htb!]
\includegraphics[width=1\columnwidth]{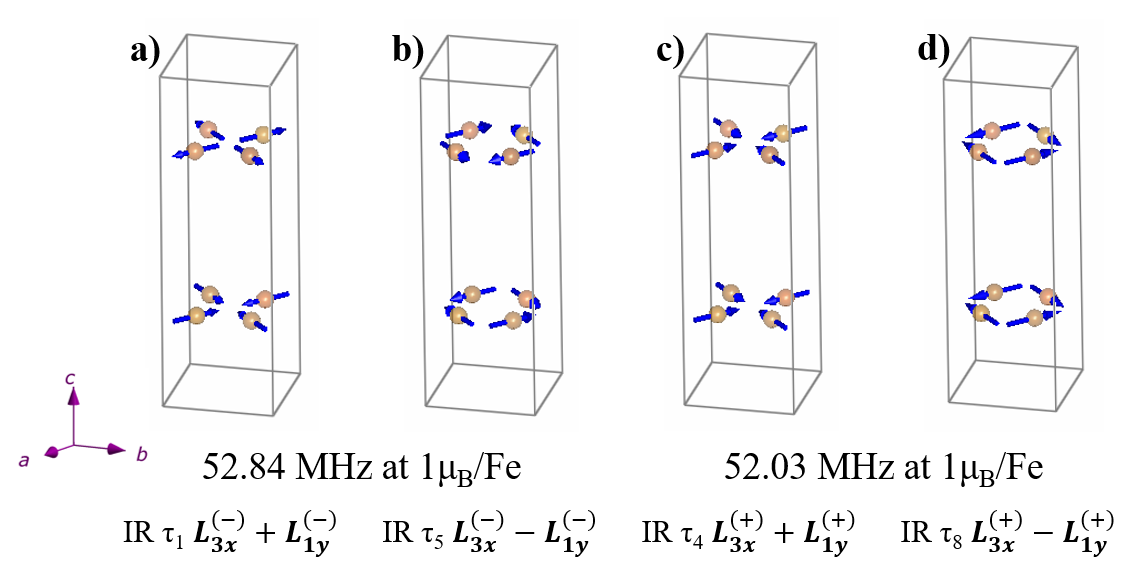}
\caption{(color online) SVC hedgehog double-\textbf{Q} structures with in-plane oriented moments in the tetragonal phase which create a finite magnetic dipole field along the \(c\)-axis at the muons sites. Shown are only the iron atoms. Panels \textbf{a)} and \textbf{b)} show the structures with AFM order along the \(c\)-axis, panels \textbf{c)} and \textbf{d)} the corresponding structures with FM order. The indicated \(\mu\)SR precession frequencies are calculated using Eqs.~\ref{eqAppB3} and~\ref{eqAppB4}. They are very similar and thus are likely within the error bar of a typical \(\mu\)SR experiment. }
\label{Fig19}
\end{figure}

\begin{figure}[htb!]
\includegraphics[width=0.5\columnwidth]{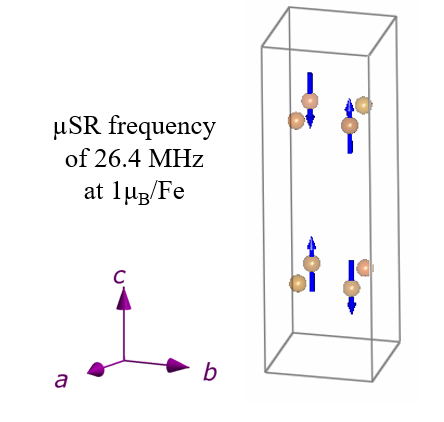}
\caption{(color online) Inhomogeneous, double-\textbf{Q} magnetic structure i-SCDW, \(L^{(-)}_{1z}+L^{(-)}_{3z}\), with alternating zero and nonzero magnetic moment at the iron sites with P\textsubscript{C}4\textsubscript{2}/ncm magnetic group symmetry according to Ref.~\cite{KhalyavinPRB2014}. The indicated \(\mu\)SR precession frequency has been calculated using Eqs.~\ref{eqAppB3} and~\ref{eqAppB4}. Shown are only the iron atoms. This is the double-\textbf{Q} magnetic structure that is compatible with our \(\mu\)SR data.}
\label{Fig20}
\end{figure}

\begin{figure}[htb!]
\includegraphics[width=0.9\columnwidth]{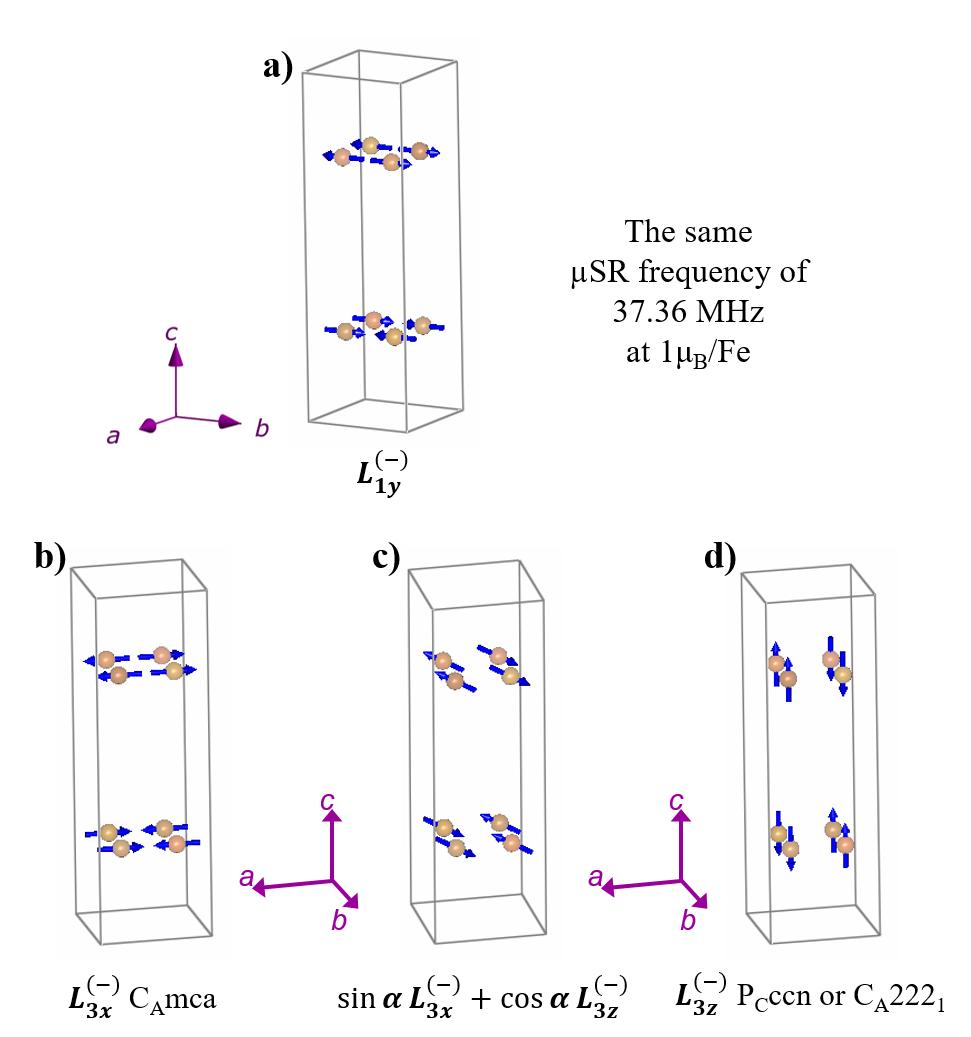}
\caption{(color online) Magnetic structures and their order parameters in the orthorhombic state. All phases preserve the same pattern (type) of the exchange interactions. \textbf{a)} and \textbf{b)} show the two domain state of the stripe-like AF order that is realized in the orthorhombic phase; \textbf{c)} a spin rotated phase with an arbitrary rotation angle \(\alpha\); \textbf{d)} the pure out-of-plane magnetic order that has been suggested in Ref.~\cite{WasserPRBR2015} as the magnetic structure in the tetragonal AF phase. This structure breaks the \(C4\) symmetry of the tetragonal crystal structure. Note that the \(\mu\)SR precession frequencies remain the same in accordance with Eqs.~\ref{eqAppB3} and \ref{eqAppB4}, under the continuous rotation from the pure in-plane to the pure out-of-plane structure. }
\label{Fig21}
\end{figure}

In Figure~\ref{Fig19} we show the non-collinear double-\textbf{Q} structures with in-plane oriented magnetic moments which create a finite dipolar magnetic field at the muons sites of tetragonal BKFA or BNFA. 
All of them belong to the hedgehog-type SVC structures. 
The indicated \(\mu\)SR precession frequencies have been obtained using Eqs.~\ref{eqAppB2} and \ref{eqAppB3} under the assumption that each Fe ion has a magnetic moment of 1~\(\mu_B\)~\cite{BernhardPRB2012, AczelPRB2008}. 
These local fields are larger than the ones calculated for the single-\textbf{Q} magnetic order in the o-AF state (see below and Fig.~\ref{Fig21}) as well as for the double-\textbf{Q} magnetic order of the tetragonal i-SCDW state (see Fig.\ref{Fig20}). 

For the double-\textbf{Q} magnetic structures shown above, each Fe ion has the same magnetic moment which is assumed to amount to 1~\(\mu_B\). 
However, there exists also the possibility of a so-called inhomogeneous double-\textbf{Q} magnetic structure for which the magnetic moment becomes zero for half of the Fe sites. 
It is described by the PC42/ncm magnetic group symmetry and preserves the \(C4\) symmetry. 
In our P4/mbm setting this structure corresponds to the linear combination \(L^{(-)}_{1z}+L^{(-)}_{3z}\).
In the I4/mmm setting, it is described by the linear combination of the order parameter \(\eta_z(k_1)+\eta_z(k_2)\) which belong to different arms of the K13-star. 
This structure is shown below in Fig.~\ref{Fig20}. 
The calculations show that it yields a moderate \(\mu\)SR precession frequency that is lower than the one in the orthorhombic phase (see below and Fig.~\ref{Fig21}) in agreement with the experimental data. 
In contrast to other magnetic phases, the coexistence of the nonmagnetic \((S=0)\) and magnetic \((S\neq 0)\) sites may indicate an alteration of the iron spin states of the neighboring ions. 
A large variation of the iron spin state is indeed not uncommon to the parent compounds of the iron superconductors for which the magnetic moment varies from the high spin state with \(S=2\) and a moment of 3.5~\(\mu_B/Fe\) in Rb\textsubscript{2}Fe\textsubscript{4}Se\textsubscript{5} to the low spin state with \(S=0\) in FeSe.

Finally we discuss the so-called single-\textbf{Q} magnetic structures which require an orthorhombic structure since they break the \(C4\) symmetry. 
From the magnetic symmetry point of view the symmetry reduction that takes place at the transition from the paramagnetic tetragonal I4/mmm1\('\) phase to the magnetic orthorhombic C\textsubscript{A}mca (or F\textsubscript{C}mm\('\)m\('\)) phase can be desscribed as a condensation of the magnetic order parameter \(\eta_{xy}(k_1)\) or \(\eta_{\bar{x}y}(k_1)\) in the I4/mmm setting.
Here two order parameters with different translation symmetry form two different orthorhombic domains. 
In our P4/mbm setting for the paramagnetic phase these two domains of the orthorhombic magnetic phase can be described as a condensation of the \(L^{(-)}_{3x}\) and \(L^{(-)}_{1y}\) order parameters, respectively. 
The structure with the out-of-plane direction of the magnetic moments in the tetragonal AF phase can be obtained by a continuous rotation of the magnetic moments in the \(ac\)-plane. 
The respective magnetic structures are shown in Fig.~\ref{Fig21}. 
Note that in accordance with Eqs.~\ref{eqAppB2} and \ref{eqAppB3} all of them give rise to the same \(\mu\)SR precession frequency which for a Fe moment of 1~\(\mu_B\) amounts to 32.3~MHz.

At last we mention the \(\mu\)SR precession frequency at the third muon site for the relevant magnetic structures under the assumption of a magnetic moment of 1~\(\mu_B/Fe\). 
In the o-AF state  (for the structure shown in Fig.~\ref{Fig21}(b)) it amounts to about 8.9 MHz; whereas in the t-AF state (for the structure shown in Fig.~\ref{Fig20}) it is reduced to about 6.5 MHz. 
Moreover, the direction of the field at this third muon site is parallel to the \(c\)-axis in the o-AF state and parallel to the \(ab\)-plane in the t-AF phase, similar to local magnetic field at the main muon site.

%
%
\begin{acknowledgments}
Work at the University of Fribourg was supported by the Schweizerische Nationalfonds (SNF) by Grant No. 200020-172611. K. W. acknowledges funding from the Alexander von Humboldt Foundation. K. W. acknowledges valuable discussions with Fr\'{e}d\'{e}ric Hardy. We thank Christof Neuruhrer and Bernard Grobety for their technical assistance in performing the EDX measurements. 
\end{acknowledgments}

%
%
\bibliographystyle{apsrev4-1}
\bibliography{biblio}

\end{document}